\documentclass[sn-mathphys,Numbered]{sn-jnl}
\usepackage{graphicx}%
\usepackage{multirow}%
\usepackage{amsmath,amssymb,amsfonts}%
\usepackage{amsthm}%
\usepackage{mathrsfs}%
\usepackage[title]{appendix}%
\usepackage{xcolor}%
\usepackage{textcomp}%
\usepackage{manyfoot}%
\usepackage{booktabs}%

\usepackage[caption=false,font=footnotesize]{subfig}
\usepackage{array}
\usepackage{hyperref}
\usepackage{soul}
\usepackage{adjustbox}
\usepackage[acronym]{glossaries}

\makeglossaries

\newacronym{pca}{PCA}{Principal Component Analysis}
\newacronym{mpeg}{MPEG}{Motion Picture Expert Group}
\newacronym{jpeg}{JPEG}{Joint Picture Expert Group}
\newacronym{vpcc}{V-PCC}{Video-based
Point Cloud Compression}
\newacronym{gpcc}{G-PCC}{Geometry-based Point Cloud Compression}
\newacronym{qoe}{QoE}{Quality of Experience}
\newacronym{mse}{MSE}{Mean Square Error}
\newacronym{psnr}{PSNR}{Peak-Signal-to-Noise-Ratio}
\newacronym{rgb}{RGB}{Red Green Blue}
\newacronym{cnn}{CNN}{Convolutional Neural Network}
\newacronym{mos}{MOS}{Mean Opinion Score}
\newacronym{pqs}{PQS}{Predicted Quality Score}
\newacronym{plcc}{PLCC}{Pearson Linear Correlation Coefficient}
\newacronym{srocc}{SROCC}{Spearman Rank Order Correlation Coefficient}
\newacronym{rmse}{RMSE}{Root Mean Square Error}

\begin{document}

\title[PointPCA: Point Cloud Objective Quality Assessment Using PCA-Based Descriptors]{PointPCA: Point Cloud Objective Quality Assessment Using PCA-Based Descriptors}



\author*[1]{\fnm{Evangelos} \sur{Alexiou}}\email{alexiou@xiaomi.com}

\author[1,2]{\fnm{Xuemei} \sur{Zhou}}\email{xuemei.zhou@cwi.nl}

\author[1]{\fnm{Irene} \sur{Viola}}\email{irene@cwi.nl}

\author[1,2]{\fnm{Pablo} \sur{Cesar}}\email{p.s.cesar@cwi.nl}


\affil[1]{\orgdiv{Distributed and Interactive Systems}, \orgname{Centrum Wiskunde en Informatica}, \orgaddress{\street{Science Park 123}, \city{Amsterdam}, \postcode{1098XG}, \country{The Netherlands}}}

\affil[2]{\orgdiv{Faculteit Elektrotechniek, Wiskunde en Informatica}, \orgname{Delft Institute of Technology}, \orgaddress{\street{Mekelweg 5}, \city{Delft}, \postcode{2628CD}, \country{The Netherlands}}}


\abstract{Point clouds denote a prominent solution for the representation of 3D photo-realistic content in immersive applications. 
Similarly to other imaging modalities, quality predictions for point cloud contents are vital for a wide range of applications, enabling trade-off optimizations between data quality and data size in every processing step from acquisition to rendering. 
In this work, we focus on use cases that consider human end-users consuming point cloud contents and, hence, we concentrate on visual quality metrics. 
In particular, we propose a set of perceptually relevant descriptors based on Principal Component Analysis (PCA) decomposition, which is applied to both geometry and texture data for full-reference point cloud quality assessment.
Statistical features are derived from these descriptors to characterize local shape and appearance properties for both a reference and a distorted point cloud. 
The extracted statistical features are subsequently compared to provide corresponding predictions of visual quality for the distorted point cloud.
As part of our method, a learning-based approach is proposed to fuse these individual predictors to a unified perceptual score.
We validate the accuracy of the individual predictors, as well as the unified quality scores obtained after regression against subjectively annotated datasets, showing that our metric outperforms state-of-the-art solutions.
Insights regarding design decisions are provided through exploratory studies, evaluating the performance of our metric under different parameter configurations, attribute domains, color spaces, and regression models.
A software implementation of the proposed metric is made available at the following link: \url{https://github.com/cwi-dis/pointpca}.}

\keywords{point cloud, quality metric, objective quality assessment, principal component analysis, descriptors}



\maketitle

\section{Introduction}
With the increasing popularity of extended reality technology and the adoption of depth-enhanced visual data in modern telecommunication and imaging systems, point clouds have emerged as a promising 3D content representation. 
However, a faithful rendition of 3D visual information using point clouds requires vast amounts of data, several orders of magnitude higher than what current transmission infrastructure can handle. 
Thus, reliable point cloud compression schemes are essential and have been a main focus of the \acrfull{mpeg}~\cite{Schwarz2018a} and \acrfull{jpeg}~\cite{Ebrahimi2016a} standardization bodies in the last few years.
As a result of these efforts, \acrshort{mpeg} has crafted two standards, namely\acrfull{vpcc}~\cite{MPEG-VPCC-standard_ISO}, 
and \acrfull{gpcc}~\cite{MPEG-GPCC-standard_ISO}, while the \acrshort{jpeg} Pleno~\cite{astola2020jpeg} Learning-based Point Cloud Coding standard~\cite{JPEG-standard_ISO} is under development.
These milestones are crucial to establish interoperability and facilitate the integration of point cloud technology in daily use cases.

Compression schemes often offer size reduction at the cost of added visual distortions. 
Moreover, point cloud contents might undergo signal deformations during processing, transmission, and/or rendering, which may have an additional negative effect on their perceptual quality.
Therefore, there is a need for mechanisms to quantify the induced visual impairments, enabling perceptually-based optimizations and ensuring the best \acrfull{qoe} for the end-users. 
Ground-truth ratings for the amount of visual impairments in a stimulus are obtained through subjective quality assessments. 
However, these procedures are time-consuming, costly, and essentially impractical for real-life applications.
Thus, objective quality methods that can automatically predict the visual quality of distorted stimuli are required.  

Two main types of characterization are commonly used to distinguish approaches for objective quality metrics for point cloud contents. One characterization comes from image and video objective quality metrics, and distinguishes between full-reference, reduced-reference, and no-reference metrics, based on their requirement for the reference content, some reference data, and no reference information at execution time, respectively.
An orthogonal characterization for point cloud quality metrics is based on the domain in which the metric is computed, differentiating them as projection-based and point-based~\cite{Alexiou2019b}. 
The former refers to 2D solutions, capturing geometric and textural distortions as reflected upon rendering on planar arrangements.
These methods commonly adopt or extend techniques that were devised for images in the past, and they are view- and rendering-dependent~\cite{Alexiou2019b}.
Conversely, point-based counterparts operate in the 3D point cloud domain and are rendering-agnostic. 
Both projection- and point-based schemes could rely on either conventional or learning-based approaches~\cite{Alexiou2022a}.
However, the latter are often treated as a separate category.

Full-reference metrics are widely used in scenarios such as rate-distortion optimization for efficient compression, in which there is a need for comparing and quantifying distortions added to a pristine reference to determine the best rate allocation. Point-based solutions are rendering-agnostic and offer better generalisation for cases in which the final rendering parameters are not known. Thus, in this work, we focus on a full-reference, point-based solution. 

Initial attempts of full-reference point-based metrics built on simple distances between individual points, whereas more recent algorithms utilize richer features that capture local patterns of geometric and textural information. 
The majority of modern point-based methods make use of small sets of geometric features, often focusing on specific surface properties, with normal vectors (e.g., ~\cite{Tian2017a,Alexiou2018a,Alexiou2020a}) and curvatures (e.g.,~\cite{Meynet2019a,Meynet2020a,Alexiou2020a}) being more widely used. 
Textural features typically rely on statistics of luminance or lightness (e.g.,~\cite{Alexiou2020a,Meynet2020a}) and occasionally chromatic components (e.g.,~\cite{Meynet2020a}), computed over spatial neighborhoods. 
Geometric and textural features are often linearly combined~\cite{Meynet2020a}, while more recently, 
paradigms of more advanced regression models, such as Random Forest~\cite{Hua2020a, Hua2021a} and Support Vector Regression~\cite{Zhang2022a} are gaining ground.
Employing learning-based frameworks to combine hand-crafted features offers the advantage of interpretability, while still leveraging machine learning to effectively map predictions from the extracted features to a single quality score.
Such methods have been successfully used in the field of image and video quality assessment, with VMAF being among the most renowned examples~\cite{li2016toward}.

In this paper, we introduce PointPCA, an objective quality metric that makes use of hand-crafted, interpretable descriptors of geometric and textural properties, based on \acrfull{pca}, in a learning-based framework for visual quality assessment of point clouds.
Subsets of the proposed geometric descriptors have already been used for urban classification~\cite{Chehata2009a},  
semantic interpretation~\cite{Weinmann2015a}, semantic segmentation~\cite{Hackel2016b}, contour detection~\cite{Hackel2016a}, and, more recently, no-reference objective quality assessment~\cite{Zhang2022a} of point cloud data.
We complement the existing literature by proposing an enriched set of \acrshort{pca}-based geometric and a novel set of \acrshort{pca}-based textural descriptors, with corresponding predictors fused through Random Forest regression to a single perceptual quality score in a full-reference design.
Our results show that PointPCA achieves high performance under all tested datasets, with substantial improvements over state-of-the-art metrics.
Exploratory studies are performed under different parameter configurations, color spaces, attribute-specific descriptors, and regression models to showcase the effectiveness and performance stability of our metric.
Our contributions can be summarized as follows:
\begin{itemize}
    \item We propose the use of statistical features computed from \acrshort{pca}-based descriptors to quantify point cloud geometric and textural distortions. 
    The descriptors are obtained per point after applying \acrshort{pca} over spatial neighborhoods, and capture local geometric and textural properties, while the statistical features estimate average and dispersion trends, promoting interpretability.  
    \item We choose the Random Forest algorithm to produce a unique perceptual quality score by fusing individual predictors obtained from the proposed statistical features in a non-linear manner.    
    We demonstrate the effect of the selected learning-based framework through comparison to other commonly used regression models. 
    Our results show high robustness under any non-linear method.
    \item We compare the performance of PointPCA to state-of-the-art metrics on a variety of datasets, showing gains in all datasets under consideration.
\end{itemize}

%

\section{Related work}
A brief description of point cloud objective quality assessment methods is provided below, after clustering them based on their operating principle. 
The interested reader may refer to~\cite{Alexiou2022a} for a more detailed overview. 

\subsection{Point-based objective quality metrics}
The point-to-point and point-to-plane~\cite{Tian2017a} denote the earliest attempts for the establishment of point-based objective quality metrics.
The former measures the Euclidean distance between point coordinates, 
while the latter relies on the projected error of distorted points across reference normal vectors. 
In both metrics, the \acrfull{mse} or the Hausdorff distance 
is applied over the individual, per-point error values, to deliver a global degradation score. 
In~\cite{Javaheri2020a}, the generalized Hausdorff distance is proposed to mitigate the sensitivity of the Hausdorff distance in outlying points, by excluding a percentage of the largest individual errors.
The geometric \acrfull{psnr}, defined in~\cite{M39966} for both metrics to account for differently scaled contents, was revised in~\cite{Javaheri2020c}
to consider the content's intrinsic or rendering resolution.
The plane-to-plane metric is described in~\cite{Alexiou2018a} and estimates the angular similarity of tangent planes, as expressed through unoriented normals. 
The point-to-distribution metric, introduced in~\cite{Javaheri2020b}, computes the Mahalanobis distance between a distorted point and a reference neighborhood. 
The PC-MSDM~\cite{Meynet2019a} evaluates the similarity of local curvature statistics, extracted after quadratic fitting in support regions. 

Previous metrics examine only geometric distortions.
A few more recent attempts employ textural-only information, albeit, the majority of metrics incorporate both geometric and textural information.
Specifically, the first texture-only metric follows the point-to-point logic and measures the \acrshort{mse} or \acrshort{psnr}~\cite{M40522}, analogously to the well-known 2D image counterpart.
More sophisticated texture-only paradigms are proposed in~\cite{Viola2020a}, which compute histograms or correlograms of luminance and chrominance components, to characterize color distributions. 

Regarding metrics that consider both geometry and texture, the point-to-distribution metric was extended to capture color degradations in~\cite{Javaheri2021a} by additionally applying the same formula on the luminance component. 
The PC-MSDM was extended to PCQM~\cite{Meynet2020a} by incorporating local statistical measurements from luminance, chrominance, and hue in order to evaluate textural impairments. 
The PointSSIM~\cite{Alexiou2020a} relies on statistical dispersion of location, normal, curvature, and luminance data. An optional pre-processing step of voxelization is proposed to enable different scaling effects and reduce intrinsic geometric resolution differences across contents.
The VQA-CPC~\cite{Hua2020a} computes statistics upon Euclidean distances between every sample and the arithmetic mean of the point cloud, using geometric coordinates and color values.
An extension is presented in~\cite{Hua2021a}, namely CPC-GSCT, which involves a point cloud partition stage, before extraction of features per region.

A graph signal processing-based approach, namely GraphSIM, is described in~\cite{Yang2020a} and evaluates statistical moments of color gradients on keypoints, after high-pass filtering on the pristine content's topology.
A multi-scale version, namely MS-GraphSIM, is presented in~\cite{ZhangY2021a}.
In~\cite{Diniz2020a}, local binary patterns are applied to the luminance component of neighboring points.
This work is extended in~\cite{Diniz2020b} considering the point-to-plane distance between point clouds, and the point-to-point distance between feature maps. 
A variant descriptor called local luminance pattern is proposed in~\cite{Diniz2020c}, introducing a voxelization stage. 
A textural descriptor to compare neighboring color values using the CIEDE2000 distance is reported in~\cite{Diniz2021a}.  
The color differences are coded as bit-based labels, which denote frequency values of pre-defined intervals.
An extension is presented in~\cite{Diniz2021b}, namely, BitDance, which incorporates bit-based labels from a geometric descriptor that relies on the comparison of neighboring normal vectors. 
The EPES presented in~\cite{Xu2022a}, relies on potential energy; that is, the energy needed to move points of a local neighborhood from an origin to their current geometric and color status. 
The MPED~\cite{yang2023a} also utilizes the point potential energy, which quantifies the spatial distribution and color under certain metric space to measure isometrical distortion. 
The potential energy discrepancy is further extended to a multiscale form.

The aforementioned are full-reference metrics.
Fewer attempts have been reported for reduced-reference and no-reference metrics. 
In particular, the first reduced-reference objective quality metric, PCM\_RR, is described in~\cite{Viola2020b} and relies on global features that are extracted from location, color, and normal data. 
More recently, a reduced-reference metric for point clouds encoded with \acrshort{vpcc} is presented in~\cite{Liu2021a}.
It is based on a linear model of geometry and color quantization parameters, with the model's parameters determined by a local and a global color fluctuation feature.
A no-reference method, namely BQE-CVP, is proposed in~\cite{Hua2021b} that combines point-based geometric features, point-based and projection-based texture degradations, and a joint geometric-color feature.
In~\cite{Zhang2022a}, the logic of using natural scene statistics for no-reference quality assessment of 2D images, is extended to 3D contents.
Specifically, the authors propose statistical properties of geometric features and LAB color value distributions, to evaluate the visual quality of both point clouds and meshes.

\subsection{Projection-based objective quality metrics}
The prediction accuracy of 2D quality metrics 
over images obtained after projecting point clouds on the six faces of a surrounding cube, was initially examined in~\cite{Torlig2018a}.
The influence of the number of viewpoints in denser camera arrangements and the exclusion of background pixels is explored in~\cite{Alexiou2019a}, which also proposes a weighting scheme based on user interactivity. 
In~\cite{Yang2020b}, a weighted combination of global and local features extracted from texture and depth images, is defined.
The Jensen-Shannon divergence on the luminance component serves as the global feature, whereas a depth-edge map, a texture similarity map, and an estimated content complexity factor account for the local features.
In~\cite{He2021a}, color and curvature values are projected on planar surfaces.
Color impairments are evaluated using probabilities of local intensity differences, 
together with statistics of their residual intensities, and similarity values between chromatic components.
Geometric distortions are assessed based on statistics of curvature residuals.
A hybrid approach using both projection- and point-based algorithms is proposed in~\cite{Chen2021a}, namely, LP-PCQM. 
The point clouds are divided into non-overlapping partitions called layers,   
with a planarization process taking place at each layer, before applying the IW-SSIM~\cite{Wang2011a} to assess geometric distortions.
Color impairments are evaluated using RGB-based variants of similarity measurements defined in~\cite{Meynet2020a}. 
In~\cite{Javaheri2022a}, an image-based metric is proposed that tackles misalignment between the original and the distorted geometry. 
This is achieved by mapping the color of the distorted point cloud to the original geometry.
The resulting and the original point clouds are then projected to the six faces of a surrounding cube, followed by cropping and padding to eliminate background pixels, before the execution of any 2D quality metric. 
The same process is repeated after mapping the original color to the distorted geometry, and a total quality score is obtained as a weighted average.

\subsection{Learning-based objective quality metrics}
In~\cite{Chetouani2021a}, \acrfull{cnn} pre-trained for classification is evaluated in the task of no-reference point cloud quality assessment, after necessary adjustments.
Geometric distances, mean curvatures, and luminance values are packed into patches, with patch quality indexes computed using a \acrshort{cnn}, and a global score obtained after pooling.
An extension of this metric for full-reference quality assessment is presented in~\cite{Chetouani2021b}.
In~\cite{Quach2021a}, the use of perceptual loss is extended to point clouds, represented as voxel grids or truncated signed distances.
The perceptual loss is applied to the latent space, after a simple auto-encoding architecture of convolution layers.
In~\cite{Liu2021b} a neural network architecture for no-reference quality assessment based on projected views is proposed, namely PQA-Net.
Features are extracted after a series of \acrshort{cnn} blocks and are shared between a distortion identifier and a quality prediction unit to obtain a final quality score.
In~\cite{Tao2021a}, the PM-BVQA is proposed, which relies on a CNN-based joint color-geometric feature extractor that is fed with corresponding projections maps, followed by a two-stage multi-scale feature fusion step, and a spatial pooling module. 
In \cite{zhang2022mm}, point clouds are split into sub-models for geometry representation and 2D image projections for texture representation, with both modalities encoded using PointNet++ and ResNet50, respectively. 
Symmetric cross-modal attention is employed to fuse multi-modality quality-aware information. 
In~\cite{shan2023gpa}, a graph convolution kernel (GPAConv) is introduced to capture the perturbation of structure and texture. 
Subsequently, the network employs a multi-task framework, with quality regression as the main task, and auxiliary tasks for predicting distortion type and degree. 
A coordinate normalization module is employed to enhance the stability of GPAConv results when confronted with shifts, scales, and rotations.

\section{Description of PointPCA}
\label{sec:pointpca}
\begin{figure*}[!t]
\centering
\includegraphics[width=0.99\textwidth]{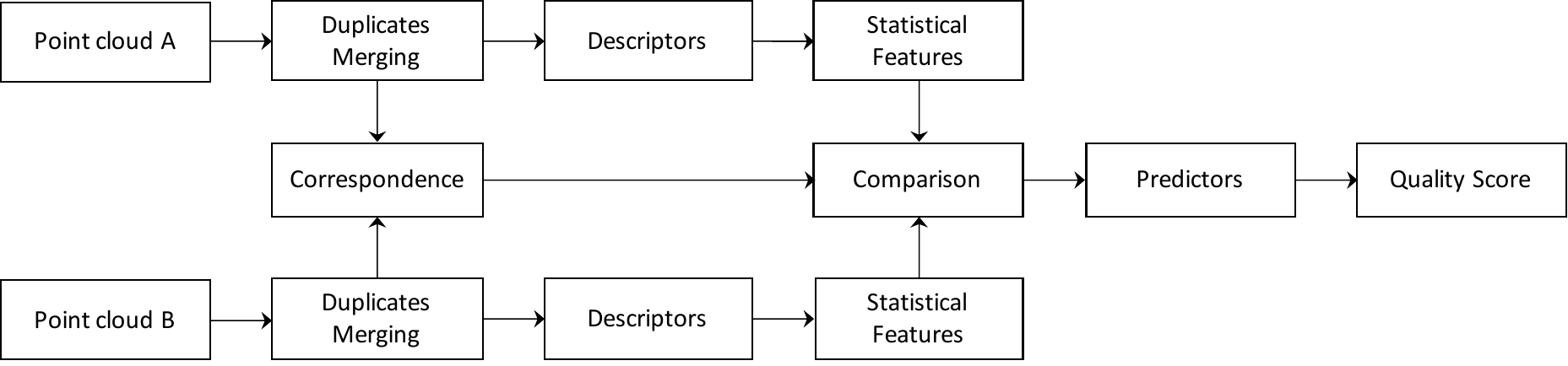}
\caption{PointPCA architecture: both the reference (i.e., Point cloud A) and the point cloud under evaluation (i.e., Point cloud B) are passing from the Duplicates Merging, computation of Descriptors, and computation of Statistical Features stages. 
After Duplicates Merging, the Correspondence between the two point clouds is computed and used for the Comparison of Statistical Features. 
A Predictor of visual quality is obtained per Statistical Feature, and all Predictors are finally fused to a total Quality Score through learning-based regression.} 
\label{fig:architecture}
\end{figure*}

The architecture of the proposed metric can be decomposed into seven stages, namely, (a)~Duplicates Merging, (b)~Correspondence, (c)~Descriptors, (d)~Statistical Features,  (e)~Comparison, (f)~Predictors, and (g)~Quality Score. 
A corresponding system diagram is presented in Figure~\ref{fig:architecture}.
The metric requires a reference during execution in order to provide a quality prediction for a point cloud under evaluation. 
Specifically, a correspondence between the two point clouds is obtained after merging points with identical coordinates that belong to the same point cloud.
Then, 23 geometric and textural descriptors are computed per point, for both point clouds. 
For every descriptor, we capture local relations by applying statistical functions, leading to corresponding statistical features. 
Given the correspondence, 46 statistical features extracted from the reference and the point cloud under evaluation per point, are compared. 
The derived error samples are pooled together, resulting in a predictor of visual quality per statistical feature.
The obtained 46 predictors are finally fused by means of a regression algorithm to obtain a total quality score for the point cloud under evaluation. 
Below, every stage is detailed separately. 

\subsection{Duplicates Merging}
Within a single point cloud, points that have identical coordinates are identified and merged; that is, only one point per coordinate set is kept~\cite{Tian2017a, Alexiou2020a}.
The color of the merged point is obtained by averaging the color of respective points with the same coordinates.
This offers the advantage that points with unique locations form neighborhoods to compute descriptors and statistical features, eliminating bias due to duplicated values. 
Moreover, redundant correspondences between a reference and a point cloud under evaluation are avoided. 

\subsection{Correspondence}
\label{ssec:correspondence}
Identifying matches between two sets of points is an ill-posed problem.
To favor lower complexity, we use the nearest neighbor algorithm for the identification of correspondences between two point clouds, similar to the majority of existing metrics (e.g.,~\cite{Tian2017a,Alexiou2018a,Alexiou2020a}).
For this purpose, one point cloud is set as the reference and the other as the point cloud under evaluation. 
Then, for every point $\mathbf{b}_i$ that belongs to the point cloud under evaluation $\mathcal{B}$ (i.e., $\mathbf{b}_{i} \in \mathcal{B}$), a matching point $\mathbf{a}_i \in \mathcal{A}$ is identified as its nearest neighbor in terms of Euclidean distance, and is registered as its correspondence.
Formally, for the point cloud under evaluation, the correspondence function is defined as $c^{\mathcal{B}, \mathcal{A}}: \mathcal{B} \xrightarrow{} \mathcal{A}$
with $c^{\mathcal{B}, \mathcal{A}}(\mathbf{b}_i) = \mathbf{a}_i$. 

Note that, different sets of matching points are obtained when iterating over the points of $\mathcal{B}$ to identify nearest neighbors in $\mathcal{A}$, with respect to starting from $\mathcal{A}$ to find matches in~$\mathcal{B}$; that is when setting $\mathcal{A}$, or $\mathcal{B}$ as reference, respectively.
In our case, we set both the pristine and the impaired point clouds as reference, as further described in Section~\ref{ssec:predictors}, and we use a max operation~\cite{Tian2017a,Alexiou2018a,Alexiou2020a} to obtain a final prediction that is independent of the reference selection. 
This is commonly referred to in the literature as symmetric error~\cite{Alexiou2022a}. 

\begin{table}[!t]
\caption{Definition of descriptors.}
\vspace{-0.5em}
\centering
\renewcommand{\arraystretch}{1.4}
\begin{tabular}{l l l}
\toprule
& Descriptor          & Definition       \\ \midrule
\multirow{13}{*}{\rotatebox[origin=c]{90}{{\centering Geometric}}}
& First eigenvalue         & $d^{g}_{1} = \lambda^{g}_{1}$ \\
& Second eigenvalue         & $d^{g}_{2} = \lambda^{g}_{2}$ \\
& Third eigenvalue         & $d^{g}_{3} = \lambda^{g}_{3}$ \\
& Sum of eigenvalues  & $d^{g}_{4} = \sum_{v=1}^{3} \lambda^{g}_v$ \\
& Linearity           & $d^{g}_{5} = (\lambda^{g}_1 - \lambda^{g}_2)/ \lambda^{g}_1$ \\
& Planarity           & $d^{g}_{6} = (\lambda^{g}_2 - \lambda^{g}_3) / \lambda^{g}_1$ \\ 
& Sphericity          & $d^{g}_{7} = \lambda^{g}_3 / \lambda^{g}_1$ \\ 
& Anisotropy          & $d^{g}_{8} = (\lambda^{g}_1 - \lambda^{g}_3) / \lambda^{g}_1$ \\
& Omnivariance        & $d^{g}_{9} = \sqrt[3]{\lambda^{g}_1 \cdot \lambda^{g}_2 \cdot \lambda^{g}_3}$ \\
& Eigenentropy        & $d^{g}_{10} = - \sum_{v=1}^{3} \lambda^{g}_v \cdot \mbox{ln}(\lambda^{g}_{v})$ \\
& Surface variation   & $d^{g}_{11} = \lambda^{g}_3 \big/ \sum_{v=1}^{3} \lambda^{g}_v$ \\
& Roughness           & $d^{g}_{12} = \left| \left( \mathbf{p}_{i} - \bar{\mathbf{p}}_{i} \right) \cdot \mathbf{e}^{g}_3 \right|$\\
& Parallelity$_x$     & $d^{g}_{13} = 1 - | \mathbf{u}_{x} \cdot \mathbf{e}^{g}_3 |$ \\
& Parallelity$_y$     & $d^{g}_{14} = 1 - | \mathbf{u}_{y} \cdot \mathbf{e}^{g}_3 |$ \\
& Parallelity$_z$     & $d^{g}_{15} = 1 - | \mathbf{u}_{z} \cdot \mathbf{e}^{g}_3 |$ \\ 
\midrule
\multirow{4}{*}{\rotatebox[origin=c]{90}{{\centering Textural}}}
& Red channel         & $d^{t}_{1} = \text{R}$ \\
& Green channel       & $d^{t}_{2} = \text{G}$ \\
& Blue channel        & $d^{t}_{3} = \text{B}$ \\
& First eigenvalue         & $d^{t}_{4} = \lambda^{t}_{1}$ \\
& Second eigenvalue         & $d^{t}_{5} = \lambda^{t}_{2}$ \\
& Third eigenvalue         & $d^{t}_{6} = \lambda^{t}_{3}$ \\
& Sum of eigenvalues  & $d^{t}_{7} = \sum_{v=1}^{3} \lambda^{t}_{v}$ \\
& Eigenentropy        & $d^{t}_{8} = - \sum_{v=1}^{3} \lambda^{t}_{v} \cdot \mbox{ln}(\lambda^{t}_{v})$ \\
 \bottomrule
\end{tabular}
\label{tbl:descDefinition}
\vspace{-1em}
\end{table}

\subsection{Descriptors}
\label{ssec:descriptors}
A set of 15 geometric and 8 textural descriptors is defined per point, to reflect local properties of point cloud topology and appearance, respectively.
The majority of those descriptors are extracted after applying \acrshort{pca} on spatial neighborhoods of geometric coordinates and textural values, correspondingly.
Specifically, provided a query point $\mathbf{p}_i$, we identify a surrounding support region that belongs to the same point cloud, forming a set $\mathbf{P}_i$ that consists of points $\mathbf{p}_{n} \in \mathbf{P}_i$.
The covariance matrix $\mathbf{\Sigma}_i$ of this set is computed, as shown in Equation~\ref{eq:cov},

\begin{equation}
\mathbf{\Sigma}_i = \frac{1}{|\mathbf{P}_{i}|} \textstyle \sum_{n = 1}^{|\mathbf{P}_{i}|} (\mathbf{p}_n - \mathbf{\bar{p}}_{i}) (\mathbf{p}_n - \mathbf{\bar{p}}_{i})^T
\label{eq:cov}
\end{equation}
with $|\mathbf{P}_{i}|$ indicating the cardinality, and $\mathbf{\bar{p}}_{i}$ the centroid of $\mathbf{P}_i$, which is given in Equation~\ref{eq:centroid}.

\begin{equation}
\mathbf{\bar{p}}_i = \frac{1}{|\mathbf{P}_{i}|} \textstyle \sum_{n = 1}^{|\mathbf{P}_{i}|} \mathbf{p}_n
\label{eq:centroid}
\end{equation}

Eigen-decomposition is then applied to the covariance matrix, 
which is symmetric and positive definite and, thus, its eigenvalues exist, are non-negative, and correspond to an orthogonal system of eigenvectors.
Eigenvectors indicate directions across which the data are mostly dispersed, while eigenvalues denote the variance of the transformed data across the principal axes.

\subsubsection*{Geometric descriptors}
For the computation of geometric descriptors, the coordinates of the points that belong in $\mathbf{P}_i$ are used; 
hence, in Equation~\ref{eq:cov} and~\ref{eq:centroid}, we set $\mathbf{p}_i = (x_i, y_i, z_i)^T$.
Let us assume that $\mathbf{e}^{g}_{1}$, $\mathbf{e}^{g}_{2}$, and $\mathbf{e}^{g}_{3}$ denote the eigenvectors that correspond to the eigenvalues $\lambda^{g}_1$, $\lambda^{g}_2$ and $\lambda^{g}_3$, with $\lambda^{g}_1 > \lambda^{g}_2 > \lambda^{g}_3$ obtained after eigen-decomposition of the covariance matrix.
Moreover, let us define $\mathbf{u}_x = (1, 0, 0)^T$, $\mathbf{u}_y = (0, 1, 0)^T$ and $\mathbf{u}_z = (0, 0, 1)^T$ to depict unit vectors across the $x$, $y$ and $z$ axis, respectively. 
Eigenvalues, eigenvectors, and unit vectors are employed to construct the proposed geometric descriptors, $\mathbf{d}^{g} \in  \mathbb{R}^{1 \times 15}$, which are defined in Table~\ref{tbl:descDefinition}.
As can be seen, each descriptor corresponds to an interpretable shape property.
Intuitively, $d^{g}_{1-4}$ denote the individual (i.e., $d^{g}_{1-3}$) and the aggregated sum (i.e., $d^{g}_{4}$) of eigenvalues that indicate dispersion magnitudes for the points distribution across the principal axes.
$d^{g}_{5-7}$ reveal behaviors of a neighborhood's points arrangement, capturing the dimensionality of the local surface.
$d^{g}_{8}$ focuses on data variation across the $1^{\text{st}}$ and the $3^{\text{rd}}$  principal directions.
$d^{g}_{9-11}$ provide an estimate of spread, uncertainty, and variation of the underlying surface, respectively, considering all principal axes.
$d^{g}_{12}$ quantifies the projected error of a queried point from its neighborhood's centroid, across the estimated normal vector, $\mathbf{e}^{g}_3$.
Finally, $d^{g}_{13-15}$ measure the projected error of $\mathbf{e}^{g}_3$ across unit vectors parallel to the Cartesian coordinate system axes where a point cloud is laying.
In summary, $d^{g}_{1-11}$ capture patterns in data dispersion, $d^{g}_{12}$ local roughness, and $d^{g}_{13-15}$ the direction of data dispersion.

\subsubsection*{Textural descriptors}
The \acrfull{rgb} color values serve as the first three descriptors of a point, noted as $d^{t}_{1-3}$.
For the computation of \acrshort{pca}-based textural descriptors, the \acrshort{rgb} color values of the points that belong in $\mathbf{P}_i$ are employed; hence, we set $\mathbf{p}_i = (\text{R}_i, \text{G}_i, \text{B}_i)^T$ in Equation~\ref{eq:cov} and~\ref{eq:centroid} and obtain the eigenvalues $\lambda^{t}_1$, $\lambda^{t}_2$ and $\lambda^{t}_3$, with $\lambda^{t}_1 > \lambda^{t}_2 > \lambda^{t}_3$. 
The individual (i.e., $d^{t}_{4-6}$) and the aggregated sum (i.e., $d^{t}_{7}$) of eigenvalues, as well as the eigenentropy (i.e., $d^{t}_{8}$) are computed to estimate dispersion magnitudes and uncertainty of the color distribution across one or all principal axes of a local neighborhood, respectively.
The formal definition of the textural descriptors, $\mathbf{d}^{t} \in  \mathbb{R}^{1 \times 8}$, is given in Table~\ref{tbl:descDefinition}.

\subsubsection*{Support regions}
A support region is required around every point sample in order to compute corresponding descriptors.
Note that for both geometric and textural \acrshort{pca}-based descriptors (i.e., all excluding $d^{t}_{1-3}$), the same support region is used and is specified based on spatial vicinity. 
In general, there are two alternatives widely employed to specify point cloud neighborhoods; that is, the $k$ nearest neighbor and the range search algorithms, hereafter, noted as $k$-nn and $r$-search, respectively.
The former leads to neighborhoods of arbitrary extent and a fixed population of points ($k$), whereas the latter identifies the same spherical volumes (of radius $r$) that enclose varying numbers of samples.

We choose the $r$-search algorithm to estimate descriptors.
This is justified by our requirement to represent properties of the same surface areas in both the reference and the distorted stimuli.
This behavior is granted by the $r$-search variant, as opposed to the $k$-nn algorithm, which is susceptible to different point densities.
For example, in the presence of down-sampling, there is no difference between the size of regions identified in the pristine and the impaired point clouds using the $r$-search. 
However, when using $k$-nn, larger regions are considered in the impaired point cloud; thus, descriptor values represent properties of underlying surfaces of different sizes. 

\begin{figure*}[!t]
\centering
\subfloat[\textit{longdress}]{\includegraphics[trim={5cm 0 5cm 0}, clip, width=0.2425\textwidth]{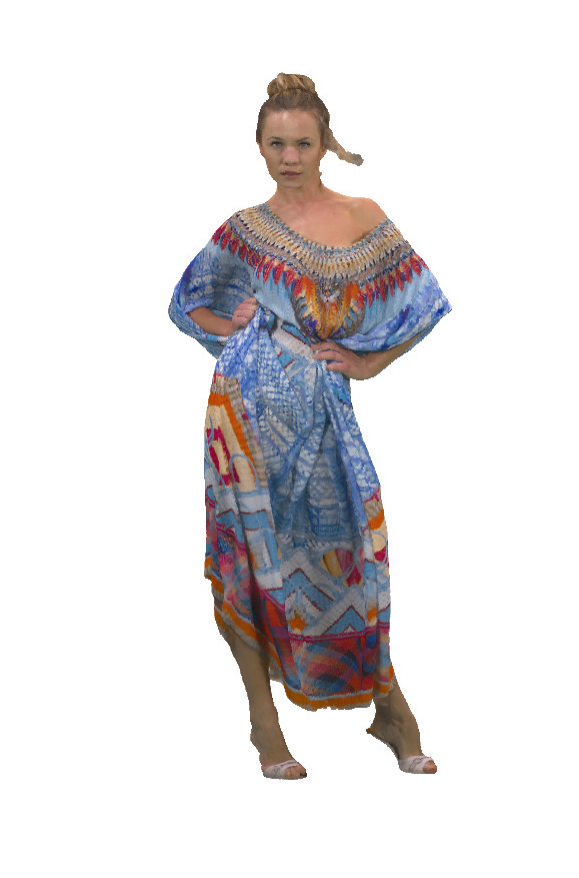}%
\label{fig:featsMap0}}
\hfil
\subfloat[${\mu}(d_{5}^{g})$]{\includegraphics[trim={5cm 0 5cm 0}, clip, width=0.2425\textwidth]{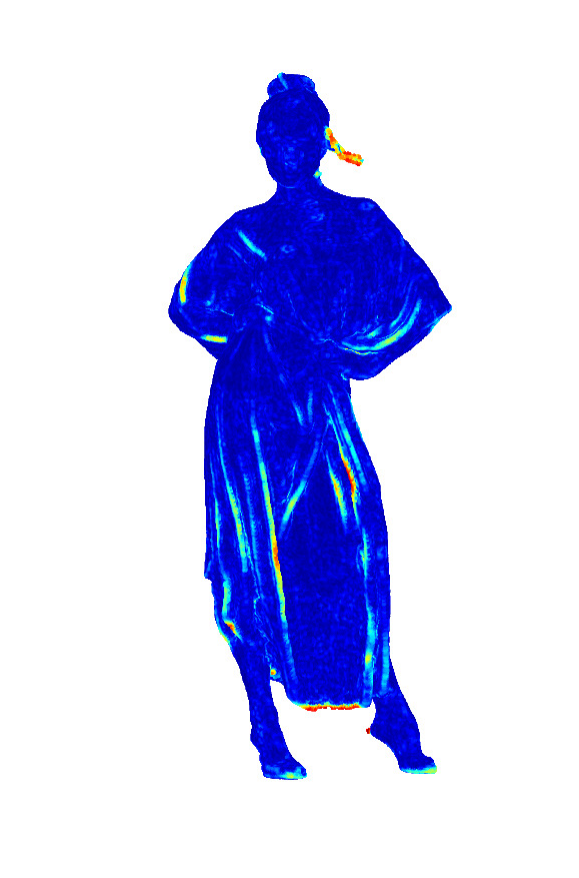}%
\label{fig:featsMap1}}
\hfil
\subfloat[${\sigma}(d_{5}^{g})$]{\includegraphics[trim={5cm 0 5cm 0}, clip, width=0.2425\textwidth]{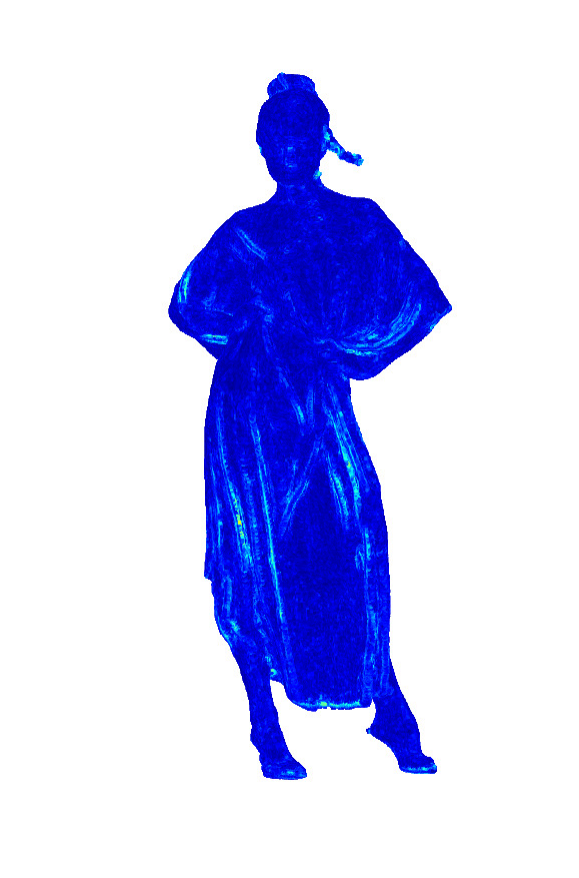}%
\label{fig:featsMap2}}
\\
\hfil
\subfloat[${\mu}(d_{6}^{g})$]{\includegraphics[trim={5cm 0 5cm 0}, clip, width=0.2425\textwidth]{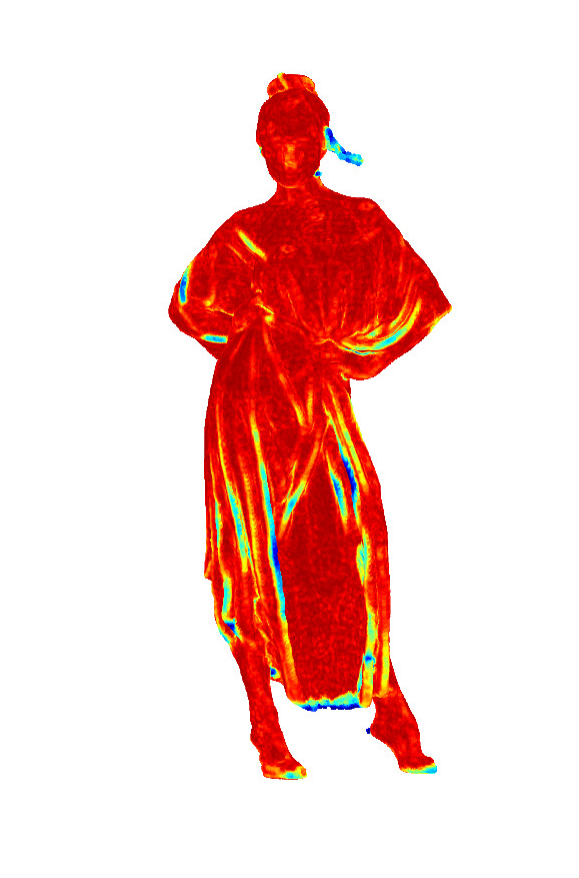}%
\label{fig:featsMap3}}
\hfil
\subfloat[${\sigma}(d_{6}^{g})$]{\includegraphics[trim={5cm 0 5cm 0}, clip, width=0.2425\textwidth]{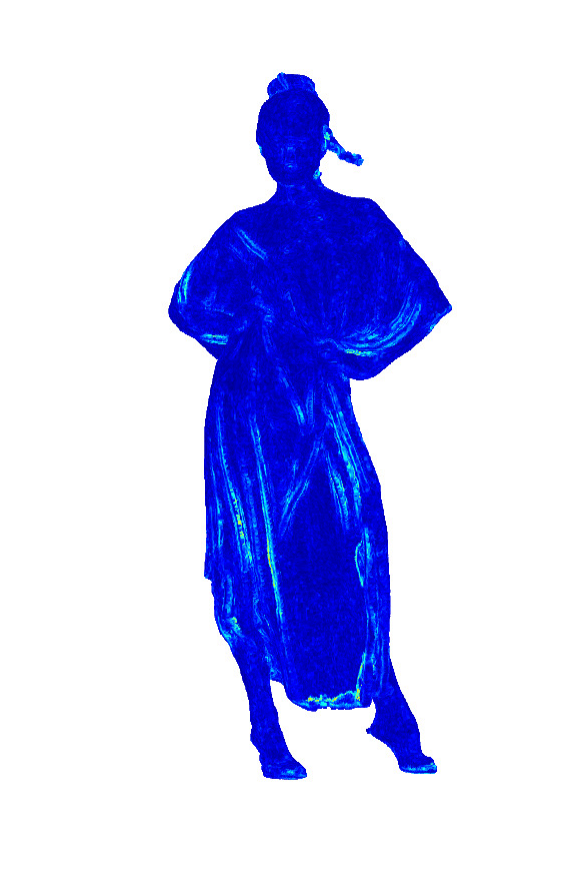}%
\label{fig:featsMap4}}
\hfil
\subfloat[${\mu}(d_{4}^{t})$]{\includegraphics[trim={5cm 0 5cm 0}, clip, width=0.2425\textwidth]{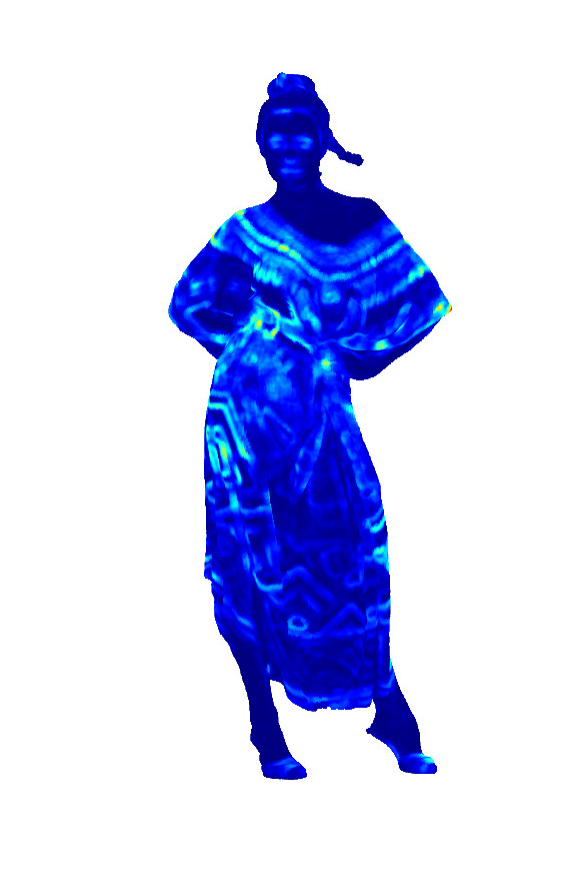}%
\label{fig:featsMap7}}
\hfil
\subfloat[${\sigma}(d_{4}^{t})$]{\includegraphics[trim={5cm 0 5cm 0}, clip, width=0.2425\textwidth]{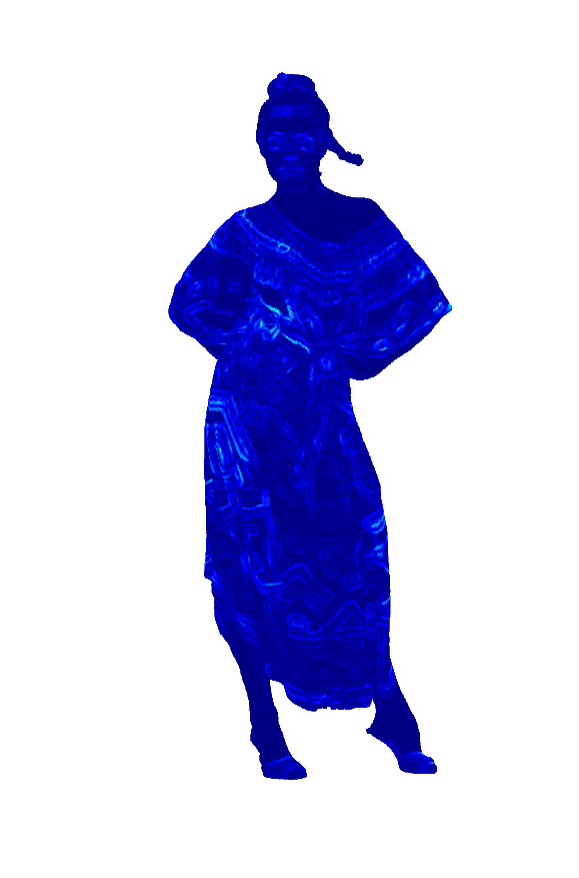}%
\label{fig:featsMap8}}
\caption{The point cloud \textit{longdress} (Figure~\ref{fig:featsMap0}) and its statistical features using the mean and standard deviation of linearity (Figures~\ref{fig:featsMap1}-\ref{fig:featsMap2}), planarity (Figures~\ref{fig:featsMap3}-\ref{fig:featsMap4}), and 
first eigenvalue on texture (Figures~\ref{fig:featsMap7}-\ref{fig:featsMap8}) descriptors. 
The amplitudes of statistical features are color-mapped, with red indicating higher and blue lower values. It can be noticed that the mean of linearity (\ref{fig:featsMap1}) and planarity (\ref{fig:featsMap3}) capture high- and low-frequency geometric regions, respectively. Moreover, the mean of the first eigenvalue on texture (\ref{fig:featsMap7}) highlights colorfulness. The standard deviation quantifies local dispersion, hence capturing high frequencies for all descriptors.}
\label{fig:featMaps}
\end{figure*}

\subsection{Statistical Features}
\label{ssec:statistical_features}
A set of 46 statistical features is computed per point, after applying 2 statistical functions to geometric and textural descriptor values that lie in the same neighborhood to capture inter-point local relations  (e.g.,~\cite{Alexiou2020a, Meynet2020a}).
In particular, the mean is computed to provide a smoother estimate of a surface property (i.e., either geometric or textural), accounting for a broader region. 
The standard deviation is also obtained, to quantify the level of variation of a surface property in the surrounding area.
Considering a query point $\mathbf{p}_i$, we identify a support region defined as a set $\mathbf{\widehat{P}}_{i}$ that consists of neighboring points $\mathbf{p}_{\hat{n}} \in \mathbf{\widehat{P}}_{i}$. 
The first statistical feature of point $\mathbf{p}_i$ is computed per Equation~\ref{eq:mean},

\begin{equation}
{\mu}_{i}(d_{u}^{\omega}) = \frac{1}{|\mathbf{\widehat{P}}_{i}|} \textstyle \sum_{\hat{n} = 1}^{|\mathbf{\widehat{P}}_{i}|} {d}_{u}^{\omega}(\mathbf{p}_{\hat{n}})
\label{eq:mean}    
\end{equation}
where $d_{u}^{\omega}(\mathbf{p}_{\hat{n}})$ denotes a descriptor relative to point $\mathbf{p}_{\hat{n}}$ from either geometry ($g$) or texture ($t$) domain $\omega \in \lbrace g, t\rbrace$, with $u \in \lbrace 1, 2, ..., 15 \rbrace$ if $\omega = g$, and $u \in \lbrace 1, 2, ..., 8 \rbrace$ if $\omega = t$. 
The second statistical feature of point $\mathbf{p}_i$ is then obtained from Equation~\ref{eq:std}.

\begin{equation}
{\sigma}_{i}(d_{u}^{\omega}) = \sqrt{\frac{1}{|\mathbf{\widehat{P}}_{i}|}  \textstyle \sum_{\hat{n} = 1}^{|\mathbf{\widehat{P}}_{i}|} \left( {d}_{u}^{\omega}(\mathbf{p}_{\hat{n}}) - {\mu}_{i}(d_{u}^{\omega}) \right)^2}
\label{eq:std}    
\end{equation}

For point $\mathbf{p}_i$, we denote with $\boldsymbol{\mu}_i \in \mathbb{R}^{1\times23}$ the concatenation of all ${\mu}_{i}(d_{u}^{\omega})$, for all descriptors from geometry followed by texture domain; analogously, we denote with $\boldsymbol{\sigma}_i \in \mathbb{R}^{1\times23}$ the concatenation of all ${\sigma}_{i}(d_{u}^{\omega})$.
A complete statistical features vector is given as $\boldsymbol{\phi}_i = [\boldsymbol{\mu}_{i}, \boldsymbol{\sigma}_{i}] \in \mathbb{R}^{1\times46}$.
In Figure~\ref{fig:featMaps}, indicative visual examples of statistical features are presented.

Statistical features are able to better capture dependencies within local neighborhoods, and provide measurements that are more perceptually coherent with respect to single points.
Specifically, they are well-aligned with primary characteristics of the human visual system, such as low-pass filtering 
and sensitivity to high-pass frequencies. 
Applying the mean in local regions mimics the former, whereas the standard deviation provides an estimate of the latter.
Moreover, statistical features are computed per point and contain contributions from its surroundings, thus, alleviating the negative effects of an erroneous correspondence, or outlying descriptor values.
That is, considering impaired stimuli that are characterized by point removal or displacement with respect to their pristine positions, errors might be introduced by the matching algorithm, or descriptor values might be poorly estimated.
Hence, comparing means instead of descriptor values mitigates the error.

\subsubsection*{Support regions}
We choose the $k$-nn algorithm to compute statistical features.
We argue that, in this case, the operating principle of this approach is beneficial for revealing topological deformations.
In particular, by appending neighboring samples until reaching $k$, we consider larger areas in a sparser impaired stimulus, and we recruit erroneous points in case of re-positioning.
Thus, larger differences will be observed in comparison to corresponding measurements taken from the pristine content.
In simpler terms, using $k$-nn allows us to penalize point sparsity and displacement. 

\subsection{Comparison}
Given the correspondence function $c^{\mathcal{B}, \mathcal{A}}(\mathbf{b}_{i}) = \mathbf{a}_i$ defined in Section~\ref{ssec:correspondence}, the $j^{\text{th}}$ statistical feature of point $\mathbf{b}_i \in \mathcal{B}$, namely $\phi^{\mathcal{B}}_{i,j}$, is compared to the $j^{\text{th}}$ statistical feature of point $\mathbf{a}_{i} \in \mathcal{A}$, namely $\phi^{\mathcal{A}}_{i,j}$ using the relative difference as in~\cite{Alexiou2020a}, per Equation~\ref{eq:relDiff},
\begin{equation}
r_{i,j}^{\mathcal{B},\mathcal{A}} = \frac{|\phi^{\mathcal{A}}_{i,j} - \phi^{\mathcal{B}}_{i,j}|}{\max \left( \left| \phi^{\mathcal{A}}_{i,j} \right|, \, \left| \phi^{\mathcal{B}}_{i,j} \right| \right) + \varepsilon}
\label{eq:relDiff}
\end{equation}
where $r_{i,j}^{\mathcal{B},\mathcal{A}}$ indicates the derived error sample that corresponds to $\mathbf{b}_i$, with $1 \leq i \leq |\mathcal{B}|$ and $1 \leq j \leq 46$, while $\varepsilon$ represents a small constant to avoid undefined operations; in this case, we use the machine rounding error for floating point numbers.
This computation is repeated for all $\mathbf{b}_i$, and corresponding error samples $r_{i,j}^{\mathcal{B},\mathcal{A}}$ are obtained. 

\subsection{Predictors}
\label{ssec:predictors}
For every statistical feature $j$, the error samples of $\mathcal{B}$ are pooled together, as shown in Equation~\ref{eq:pool}.
\begin{equation}
s_{j}^{\mathcal{B},\mathcal{A}} = \frac{1}{|\mathcal{B}|} \textstyle \sum_{i = 1}^{|\mathcal{B}|} r_{i,j}^{\mathcal{B},\mathcal{A}}
\label{eq:pool}
\end{equation}

The same computations are repeated after setting the point cloud $\mathcal{B}$ as the reference, provided the correspondence function $c^{\mathcal{A}, \mathcal{B}}(\mathbf{a}_k) = \mathbf{b}_k$, and a corresponding measurement $s_{j}^{\mathcal{A},\mathcal{B}}$ is computed.
Finally, for every statistical feature $j$, a corresponding predictor $s_j$, with $1 \le j \le 46$, is obtained after applying the symmetric max operation similarly to~\cite{Tian2017a,Alexiou2018a,Alexiou2020a}, per Equation~\ref{eq:sym}.
\begin{equation}
    s_{j} = \mathrm{max}\left( s_{j}^{\mathcal{B},\mathcal{A}}, \, s_{j}^{\mathcal{A},\mathcal{B}} \right)
    \label{eq:sym}
\end{equation}

\subsection{Quality Score}
\label{sec:quality_score}
Each predictor $s_{j}$ provides a quality rating based on the $j^{\text{th}}$ statistical feature. 
To combine all 46 predictors into a total quality score, $q$, any linear or non-linear regression model can be used.
Machine learning-based regression models have been extensively used to tackle this problem in the domain of quality assessment. 
As part of our metric, we use the Random Forest algorithm. 
This is an ensemble learning method that can improve the prediction performance with respect to single features while limiting overfitting issues.
Note that we evaluate the impact of using different regression models on the performance of our method in section~\ref{ssec:regression_models}.

\subsection{Complexity}
The total complexity of the algorithm is dominated by the operations that require the definition of a support region using the $r$-search and $k$-NN algorithms for the computation of descriptors and statistical features, as described in Sections~\ref{ssec:descriptors} and~\ref{ssec:statistical_features}, respectively.
For a given point cloud $\mathcal{P}$, such operations generally have average complexity $O(|\mathcal{P}|\log |\mathcal{P}|)$ for well-behaved cases, and $O(|\mathcal{P}|^2)$ in the worst-case scenario. Thus, an upper bound of the complexity of the algorithm can be defined as $O(N^2)$, in which $N = \max(|\mathcal{A}|, |\mathcal{B}|)$.



\section{Benchmarking setup}
\subsection{Selection of datasets}
\label{ssec:datasets}
Three subjectively annotated data sets are used to evaluate the performance of the proposed and state-of-the-art quality metrics under consideration, namely, M-PCCD (D1)~\cite{Alexiou2019b}, SJTU (D2)~\cite{Yang2020b} and WPC (D3)~\cite{Liu2022a}. 
D1 consists of 8 colored static point clouds illustrating both human figures and inanimate objects, whose geometry and color are encoded using \acrshort{vpcc} and four \acrshort{gpcc} variants (i.e., Octree-plus-Lifting, Octree-plus-RAHT, TriSoup-plus-Lifting, and TriSoup-plus-RAHT), resulting in 232 distorted stimuli. 
D2 comprises 9 colored point clouds depicting both human figures and inanimate objects that are subject to octree-based compression, color noise, geometry Gaussian noise, down-scaling, and a superposition of every combination of two aforementioned degradations excluding compression, for a sum of 378 distorted stimuli.
Finally, D3 contains 20 colored point clouds depicting inanimate objects, that are subject to octree-based down-sampling, a superposition of geometric and color Gaussian noise, and a superposition of geometric and color compression distortions using a TriSoup- and an Octree-based \acrshort{gpcc} variant, as well as \acrshort{vpcc}, for a total of 740 distorted stimuli.

\subsection{Computation of performance indexes}
To evaluate the performance of an objective quality metric in predicting perceptual quality, \acrfull{mos} from subjects participating in dedicated experiments are employed as ground truth.
The metrics are typically benchmarked after applying a fitting function to map the objective scores to the subjective quality range, while also accounting for biases, non-linearities, and saturations from subjective testing. 
Let us define a score obtained by the execution of an objective metric as a \acrfull{pqs}. 
A predicted MOS, denoted as P(MOS), is estimated by applying the fitting function on the [PQS, MOS] data set. 
In our analysis, the Recommendation ITU-T J.149~\cite{ITUTJ149} is followed, using the logistic function type~II.
Then, the \acrfull{plcc}, the \acrfull{srocc}, and the \acrfull{rmse} are computed between the P(MOS) and MOS to draw conclusions on the linearity, monotonicity, and accuracy of the objective quality metrics, respectively.

\subsection{Configuration and execution of objective quality metrics}
\label{ssec:execution}
State-of-the-art objective quality metrics are employed in our performance evaluation analysis for comparison purposes. 
In particular, we use the point-to-point, point-to-plane~\cite{Tian2017a}, and color \acrshort{psnr} on luminance component, which are being used in the \acrshort{mpeg} standardization activities for point cloud compression.
We also use the plane-to-plane~\cite{Alexiou2018a}, the joint point-to-distribution metric~\cite{Javaheri2021a} with logarithmic values, the BitDance~\cite{Diniz2021b}, the PointSSIM~\cite{Alexiou2020a} (on geometry, normal, curvature and luminance), the PCQM~\cite{Meynet2020a}, and the MPED~\cite{yang2023a}.

To compute the point-to-point and point-to-plane, the software version 0.13.5~\cite{M40522} is used. 
For the latter, the normals are computed using a quadratic fitting with $r$-search and $r = 0.01 \times B_{R}$, where $B_{R}$ indicates the maximum length of the bounding box of the reference point cloud.
For plane-to-plane, the normals are computed based on quadratic fitting with $r$-search and $r = 0.02 \times B_{R}$, following literature best practices~\cite{alexiou2020benchmarking}. 
In the point-to-distribution metric, neighborhoods consisting of $k = 31$ point samples are considered. 
For BitDance, we use the recommended configurations, namely, $k = 6$ for the target voxel edge size, while the neighborhood size is set to 6/12 and the label bits to 16/8 for geometry/color histogram.
For PointSSIM, the default parameters are employed, with the variance as the selected estimator of statistical dispersion, and $k = 12$; for the computation of curvatures and normals, quadric fitting with $r$-search and $r = 0.01 \times B_{R}$ was used.
In PCQM, the default configurations are used. 
For MPED~\cite{yang2023a}, the default settings are employed, with $L$ defined as a fraction of the total number (i.e., 1/10000), and the square of $\ell^2$ norm adopted as the distance function.
For PointPCA, the \acrshort{pca}-based descriptors (i.e., all except $d_{1-3}^{t}$) are estimated using the $r$-search with $r = 0.008 \times B_{R}$, while for the statistical features, the $k$-nn algorithm with $k = 9$ is used. 
The Random Forest regression method is implemented using the scikit-learn python framework~\cite{scikit2011} with the default configuration, namely, \acrshort{mse} as a criterion for split and 100 trees.
Note that results from the \acrshort{psnr} versions of point-to-point and point-to-plane are not reported, due to the presence of infinity values, which prevented correlation computations and fair comparison.

\subsection{Evaluation of objective quality metrics}
\label{ssec:validation}
As part of our analysis, we evaluate the performance of each individual predictor on the datasets D1, D2, and D3; in this case, all contents of each dataset are considered.
Moreover, we evaluate the performance of PointPCA after fusing individual predictors using learning-based regression. 
However, such a validation requires splitting the datasets into training and testing sets.
In our analysis, performance indexes are computed and provided only for the testing counterparts. 
In particular, PointPCA quality prediction models obtained using either Random Forest (i.e., as part of our architecture with results reported in section~\ref{ssec:performance_pointpca}) or other regression models (i.e., as part of our comparative analysis in section~\ref{ssec:regression_models}), are validated both within and across datasets using the leave-$p$-out method.
Specifically, each dataset is split into two partitions that contain 80\% and 20\% of the contents for training and testing, respectively, with all the distorted versions of a specific content placed in one partition.
For D1, D2, and D3, we use 6/2, 7/2, and 16/4 contents for training/testing, respectively. 
Then, a quality prediction model is trained on the training data and tested on the corresponding testing data of the same dataset, for within-dataset validation.
Moreover, the same quality prediction model is tested on each of the other two (entire) datasets for cross-dataset validation.
This process is repeated for all possible 80\%-20\% splits of each dataset, leading to 28, 36, and 4845 testing partitions and an equal number of corresponding quality prediction models for D1, D2, and D3 respectively. 
The average and the standard deviation of the performance indexes across all testing partitions are reported for the within-dataset validation, while only the average is reported for the cross-dataset validation.

Finally, we compare PointPCA with state-of-the-art metrics.
To enable a fair comparison between PointPCA quality models and non-learning-based metrics from the literature, performance indexes for the latter are computed over the same testing partitions.
That is, on the same testing data obtained after applying the leave-$p$-out method with 80\%-20\% splits on each dataset, separately.
Then, the average and the standard deviation of every performance index are computed across all testing partitions of each dataset (i.e., 28, 36, and 4845 testing partitions for D1, D2, and D3, respectively).

\begin{figure*}[!t]
\centering
\subfloat[D1]{\includegraphics[width=0.49\textwidth]{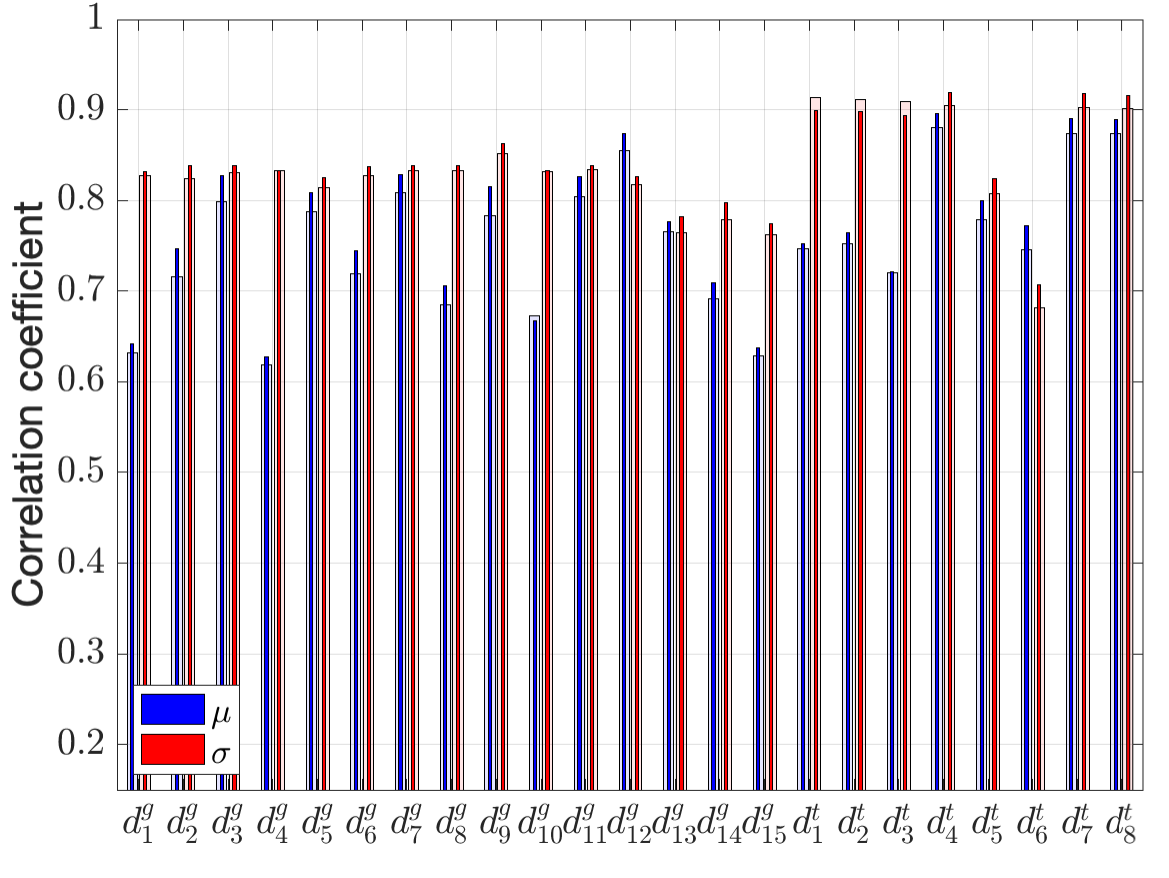}
\label{fig:predPerfD1}}
\hfil
\subfloat[D2]{\includegraphics[width=0.49\textwidth]{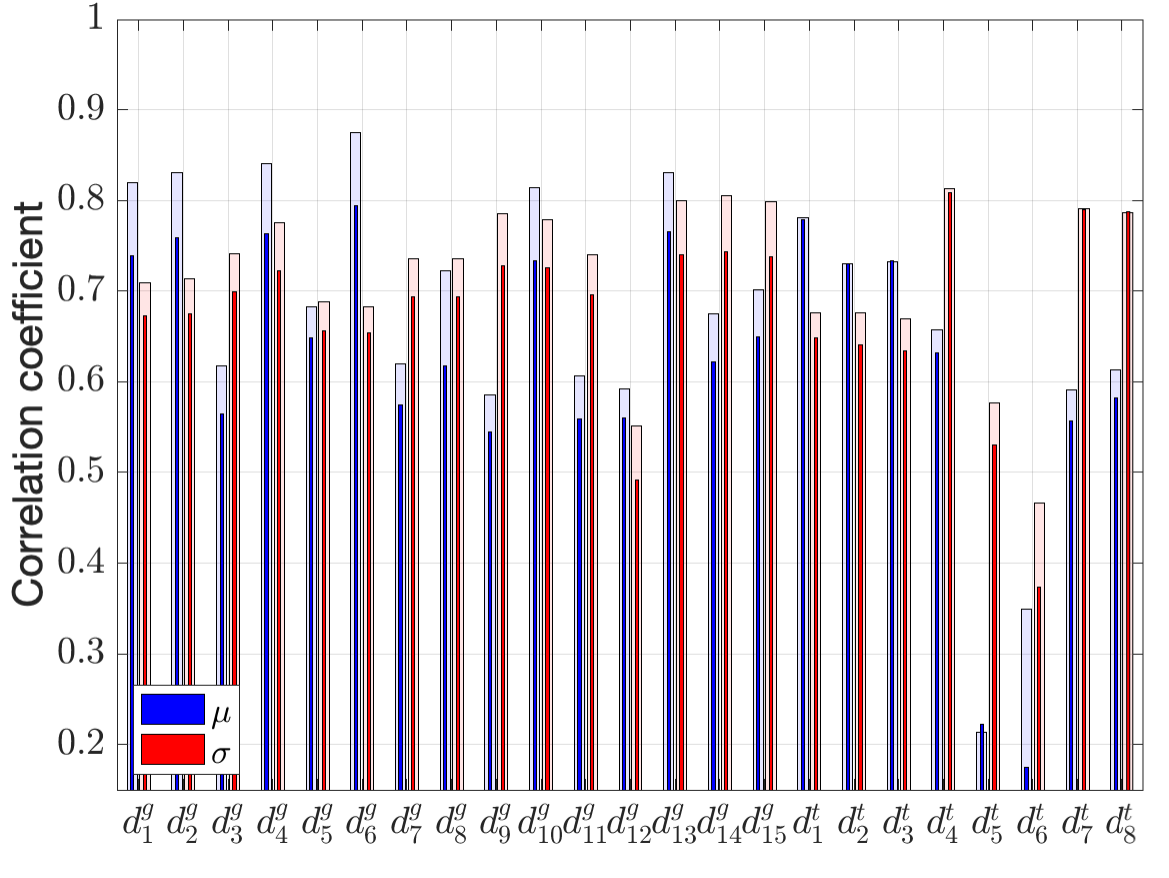}
\label{fig:predPerfD2}}
\hfil
\subfloat[D3]{\includegraphics[width=0.49\textwidth]{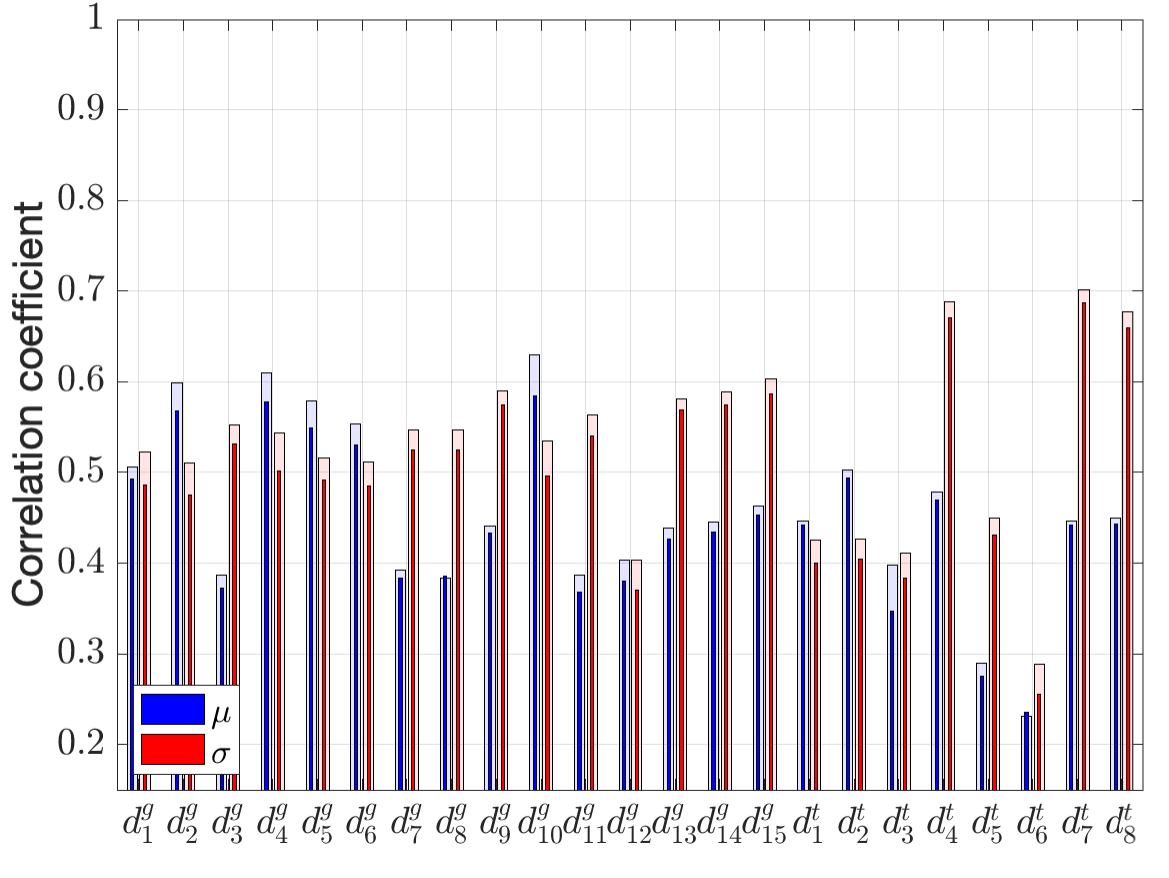}
\label{fig:predPerfD3}}
\caption{Benchmarking of predictors by means of PLCC (thin opaque bars) and SROCC (thick transparent bars), grouped per descriptor $d_{u}^{\omega}$. Each bar represents a predictor, which relies on either the ${\mu}(d_{u}^{\omega})$ or the ${\sigma}(d_{u}^{\omega})$ statistical feature, and is indicated with blue and red color, respectively.}
\label{fig:predPerf}
\end{figure*}

\section{Results}
\label{sec:results}
\subsection{Performance evaluation of predictors}
In Figure~\ref{fig:predPerf}, the \acrshort{plcc} and \acrshort{srocc} of every predictor are illustrated in the form of bars grouped per descriptor, against subjectively annotated datasets.
It can be noticed that the prediction accuracy of the proposed predictors is reaching a different performance plateau per dataset; in particular, we observe high performance for D1 and D2, whilst substantially lower for D3.
This can be explained by the different distortion characteristics of each dataset.
Specifically, geometric-only and textural-only predictors cannot accurately capture combinations of different geometric and textural degradation levels (e.g., D3),  whereas better trends are expected when the level of degradation in both geometry and texture is amplified simultaneously (e.g., D1 and D2).

Moreover, the standard deviation is found to perform better than the mean across all datasets, showing a certain level of consistency.
Specifically, for $d^{g}_{3, 7, 8, 9, 11, 14, 15}$ and $d^{t}_{4, 5, 7, 8}$ the standard deviation performs steadily better than the mean, while the mean is superior only for $d^{g}_{12}$.
For the remaining descriptors, different behaviors are observed across datasets, although the differences are limited.
For instance, for $d^{g}_{1}$, the standard deviation exhibits higher accuracy in D1 compared to the mean, with the opposite being true for D2, while equivalent performance is observed in D3.

Finally,  it is remarked that predictors using the textural descriptors $d^{t}_{4, 7, 8}$ are ranked among the best places consistently across all datasets. 
In general, they are found to be superior to every geometric predictor in D1 and D3, while in D2 they show high predictive power, despite the fact that geometric predictors perform overall better in this dataset.
The high effectiveness of textural predictors can be justified by considering that they incorporate a spatial dimension through the usage of geometric neighborhoods for the computation of descriptors and statistical features.
Therefore, they not only explicitly evaluate textural distortions, but they additionally capture topological deformations in an implicit manner.

\begin{figure}[!h]
\centering
\includegraphics[width=0.65\textwidth]{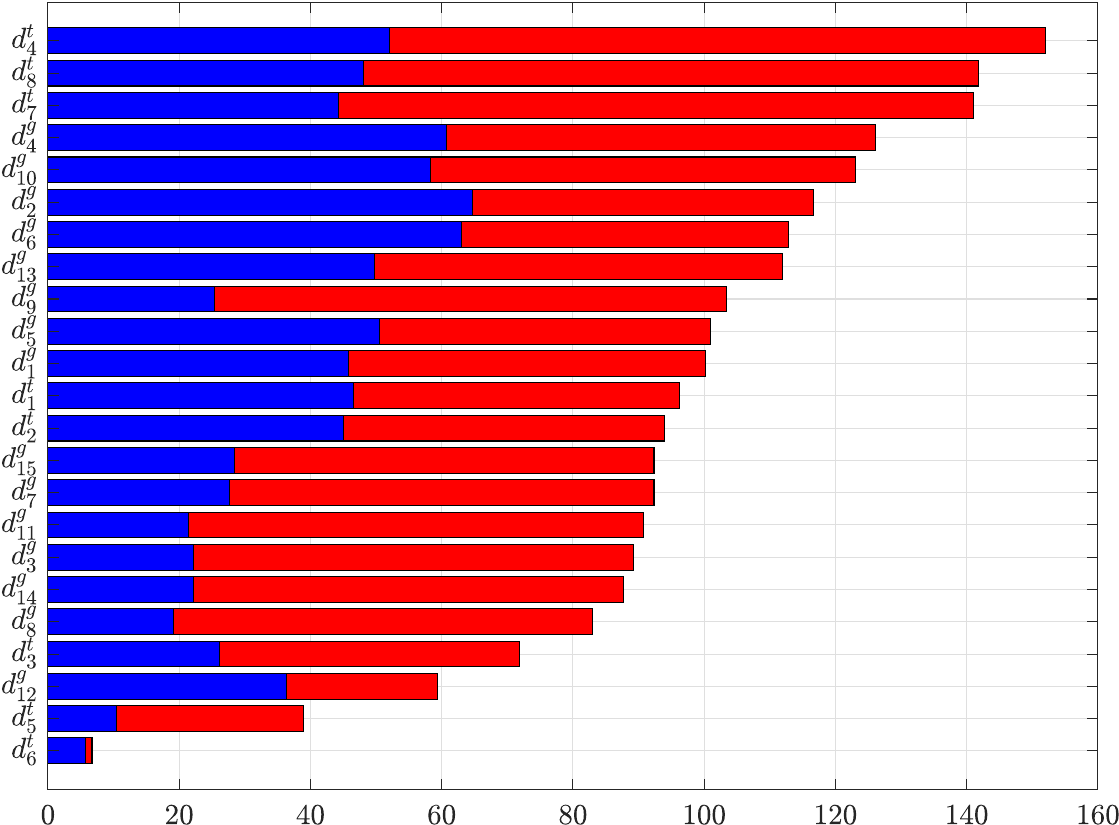}%
\caption{Importance ranking scores of predictors, computed based on their average ranking order across all datasets, stacked per descriptor $d_{u}^{\omega}$. The ranking order is determined using both PLCC and SROCC. Blue and red bars represent predictors that rely on ${\mu}(d_{u}^{\omega})$ and ${\sigma}(d_{u}^{\omega})$, respectively.}
\label{fig:predRank}
\end{figure}

The above observations are in alignment with the results presented in Figure~\ref{fig:predRank}, where the importance ranking scores of the proposed predictors are depicted.
Specifically, the average ranking order of every predictor is computed across all datasets based on the average \acrshort{plcc} and \acrshort{srocc}.
The average ranking order is then scaled to the range [1 - 100], with 1 indicating the minimum and 100 the maximum importance score, which corresponds to the lowest and highest average ranking order, respectively.
Importance ranking scores are grouped and stacked per descriptor (blue corresponds to ${\mu}(d_{u}^{\omega})$ and red to ${\sigma}(d_{u}^{\omega})$ statistical feature), before being sorted in descending order, based on their aggregated sum. Thus, the final ranking scale would range between [3 - 199].
The results show that the predictor based on $\sigma(d^{t}_{4})$ achieves the highest score, with predictors based on $\sigma(d^{t}_{7})$ and $\sigma(d^{t}_{8})$ closely following.
These results confirm the superiority of textural predictors based on $d^{t}_{4, 7, 8}$ as already noted in Figure~\ref{fig:predPerf}.

\begin{table*}[!t]
\caption{Performance evaluation of PointPCA over datasets D1, D2, and D3. Italics indicate within-dataset results. 
}
\vspace{-0.5em}
\resizebox{1\textwidth}{!}{
\label{tbl:crossdataset}
\centering
\renewcommand{\arraystretch}{1.2}
\begin{tabular}{ c c c c| c c c| c c c}
\toprule
                    & \multicolumn{9}{c}{Test} \\ \cmidrule{2-10}
& \multicolumn{3}{c|}{D1}& \multicolumn{3}{c|}{D2}  & \multicolumn{3}{c}{D3}  \\ 
 Train  & PLCC~$\uparrow$  & SROCC~$\uparrow$ & RMSE~$\downarrow$ & PLCC~$\uparrow$  & SROCC~$\uparrow$ & RMSE~$\downarrow$ & PLCC~$\uparrow$  & SROCC~$\uparrow$ & RMSE~$\downarrow$ \\ 
\midrule
D1    & \textit{0.938}	& \textit{0.942}	& \textit{0.444}	& 0.808	& 0.803	& 1.423	& 0.567	& 0.569	& 18.864 \\
D2    & 0.824	& 0.836	& 0.769	& \textit{0.932}	& \textit{0.907}	& \textit{0.859}	& 0.683	& 0.678	& 16.727 \\
D3    & 0.786	& 0.817	& 0.838	& 0.862	& 0.842	& 1.229&  \textit{0.894}	& \textit{0.890}	& \textit{10.132} \\
\bottomrule
\end{tabular}%
 }
\end{table*}

\subsection{Performance evaluation of PointPCA}
\label{ssec:performance_pointpca}
Table~\ref{tbl:crossdataset} shows the performance of PointPCA over the three selected datasets, for both within- and cross-dataset validation as described in Section~\ref{ssec:validation}. 
Substantial improvements are remarked when combining predictors with respect to using them singularly, as depicted in Figure~\ref{fig:predPerf}.
In particular, significant performance boosts are observed for D3, which is the most populated dataset with the most diverse distortion types. 
Notable gains are also shown for D2, while smaller differences are noticed for D1.

\begin{table*}[!t]
\caption{Performance evaluation of state-of-the-art quality metrics. Bold represents best and underlined second-best performance.}
\vspace{-0.5em}
\resizebox{1\textwidth}{!}{
\label{tbl:sota}
\centering
\renewcommand{\arraystretch}{1.2}
\begin{tabular}{l c c c| c c c| c c c}
\toprule
& \multicolumn{3}{c|}{D1}& \multicolumn{3}{c|}{D2}  & \multicolumn{3}{c}{D3}  \\ 
Metric                      & PLCC~$\uparrow$  & SROCC~$\uparrow$ & RMSE~$\downarrow$  & PLCC~$\uparrow$  & SROCC~$\uparrow$ & RMSE~$\downarrow$  & PLCC~$\uparrow$  & SROCC~$\uparrow$ & RMSE~$\downarrow$  \\ 
\midrule
Point-to-point\_MSE~\cite{Tian2017a}	        & 0.884 $\pm$ 0.047	& 0.896 $\pm$ 0.042	& 0.615 $\pm$ 0.142	& 0.697 $\pm$ 0.138	& 0.612 $\pm$ 0.157	& 1.662 $\pm$ 0.396	& 0.582 $\pm$ 0.067	& 0.563	$\pm$ 0.071	& 18.490 $\pm$ 0.939 \\
Point-to-plane\_MSE~\cite{Tian2017a}	        & 0.891 $\pm$ 0.033	& 0.901 $\pm$ 0.025	& 0.604 $\pm$ 0.116	& 0.663	$\pm$ 0.128	& 0.578	$\pm$ 0.155	& 1.762 $\pm$ 0.334	& 0.465	$\pm$ 0.065	& 0.452	$\pm$ 0.065	& 20.170 $\pm$ 0.696 \\
PSNR Y~\cite{M40522}	                       & 0.796 $\pm$ 0.140	& 0.798 $\pm$ 0.162	& 0.751 $\pm$ 0.270	& 0.751 $\pm$ 0.080	& 0.743 $\pm$ 0.083	& 1.561 $\pm$ 0.210	& 0.638 $\pm$ 0.059	& 0.614 $\pm$ 0.061	& 17.507 $\pm$ 1.145 \\
Plane-to-plane~\cite{Alexiou2018a}	           & 0.837 $\pm$ 0.072	& 0.847 $\pm$ 0.076	& 0.719 $\pm$ 0.161	& 0.836 $\pm$ 0.033	& 0.761 $\pm$ 0.039	& 1.311 $\pm$ 0.136	& 0.474 $\pm$ 0.068	& 0.454 $\pm$ 0.069	& 20.051 $\pm$ 0.780 \\
LOG point-to-distribution-joint~\cite{Javaheri2021a}  & 0.919 $\pm$ 0.027	& 0.921 $\pm$ 0.024	& 0.526 $\pm$ 0.106	& 0.692 $\pm$ 0.113	& 0.645 $\pm$ 0.120	& 1.698 $\pm$ 0.301	& 0.475 $\pm$ 0.079	& 0.435 $\pm$ 0.065	& 20.005 $\pm$ 0.860 \\
BitDance~\cite{Diniz2021b}	                   & 0.850	$\pm$ 0.073	& 0.859	$\pm$ 0.061	& 0.688	$\pm$ 0.161	& 0.767	$\pm$ 0.074	& 0.748	$\pm$ 0.077	& 1.521	$\pm$ 0.228	& 0.488	$\pm$ 0.055	& 0.451 $\pm$ 0.054	& 19.912 $\pm$ 0.799 \\
PointSSIM (geometry)~\cite{Alexiou2020a}	   & 0.848 $\pm$ 0.060	& 0.841 $\pm$ 0.066	& 0.700 $\pm$ 0.161	& 0.754 $\pm$ 0.039	& 0.712 $\pm$ 0.032	& 1.573 $\pm$ 0.132	& 0.402 $\pm$ 0.039	& 0.345 $\pm$ 0.054	& 20.914 $\pm$ 0.483 \\
PointSSIM (normal)~\cite{Alexiou2020a}	       & 0.831 $\pm$ 0.082	& 0.831 $\pm$ 0.082	& 0.723 $\pm$ 0.175	& 0.840 $\pm$ 0.049	& 0.764 $\pm$ 0.059	& 1.294 $\pm$ 0.227	& 0.631 $\pm$ 0.051	& 0.605 $\pm$ 0.059	& 17.664 $\pm$ 0.848 \\
PointSSIM (curvature)~\cite{Alexiou2020a}	   & 0.873 $\pm$ 0.059	& 0.876 $\pm$ 0.050	& 0.640 $\pm$ 0.152	& 0.852 $\pm$ 0.046	& 0.772 $\pm$ 0.055	& 1.250 $\pm$ 0.213	& 0.597 $\pm$ 0.056	& 0.573 $\pm$ 0.058	& 18.271 $\pm$ 0.904 \\
PointSSIM (luminance)~\cite{Alexiou2020a}	   & \underline{0.937} $\pm$ 0.026	& 0.925 $\pm$ 0.024	& \underline{0.458} $\pm$ 0.071	& 0.735 $\pm$ 0.055	& 0.708 $\pm$ 0.070	& 1.621 $\pm$ 0.181	& 0.485 $\pm$ 0.058	& 0.465 $\pm$ 0.059	& 19.945 $\pm$ 0.777 \\
PCQM~\cite{Meynet2020a}	                       & 0.930 $\pm$ 0.041	& \underline{0.940} $\pm$ 0.032	& 0.472 $\pm$ 0.136	& \underline{0.878} $\pm$ 0.025	& \underline{0.862} $\pm$ 0.030	& \underline{1.145} $\pm$ 0.121	& \underline{0.754} $\pm$ 0.034	& \underline{0.749} $\pm$ 0.036	& \underline{14.960} $\pm$ 0.892 \\
MPED~\cite{yang2023a}	                       & 0.857 $\pm$ 0.095	& 0.870 $\pm$ 0.091	& 0.657 $\pm$ 0.233	& 0.872 $\pm$ 0.056	& 0.856 $\pm$ 0.067	& 1.148 $\pm$ 0.239	& 0.618 $\pm$ 0.056	& 0.620 $\pm$ 0.055	& 17.893 $\pm$ 0.926 \\
\midrule
PointPCA                    & \textbf{0.938} $\pm$ 0.039	    & \textbf{0.942} $\pm$ 0.042	& \textbf{0.444}	$\pm$ 0.127 & \textbf{0.932}	$\pm$ 0.025	& \textbf{0.907}	$\pm$ 0.043	& \textbf{0.859}	$\pm$ 0.166 & \textbf{0.894}	$\pm$ 0.027	& \textbf{0.890}	$\pm$ 0.029	& \textbf{10.132}	$\pm$ 1.195 \\
\bottomrule
\end{tabular}
}
\end{table*}

As expected, within-dataset results generally achieve better performance with respect to cross-dataset results.
Considering cross-dataset validation results, training on D1 leads to poor generalization capabilities on D2 and D3, compared to training on D3 and D2, respectively. 
Training on D2 leads to better generalization on D1 with respect to training on D3, while the performance on D3 remains low.
These results can be explained by the intrinsic characteristics of the datasets; D1 contains only compression distortions with both human and object models, D2 additionally employs geometric and color noise, while D3 is the most diverse in terms of distortion types containing only objects (see section~\ref{ssec:datasets}).

\subsection{Comparison with the state of the art}
In Table~\ref{tbl:sota}, we show performance results of PointPCA and existing point cloud quality metrics across the selected datasets, for comparison purposes.
Specifically, we report the performance indexes as obtained from the within-dataset validation of PointPCA and the evaluation of the alternative metrics on the same testing partitions, as described in section~\ref{ssec:validation}.
Our results suggest that the PointPCA metric achieves the best performance in all datasets with high scores.
Considering D1, the luminance-based PointSSIM variant achieves the second-best performance in terms of \acrshort{plcc} and \acrshort{rmse} followed by PCQM, which attains the second-best performance in terms of \acrshort{srocc}.
The PCQM is consistently ranked as the second-best option in D2 and D3, followed by the MPED, and the normal- and curvature-based variants of the PointSSIM.
It is evident that in D2 and D3, our proposed metric achieves substantial gains in terms of \acrshort{plcc}, \acrshort{srocc}, and \acrshort{rmse} with respect to alternative metrics. 

\section{Exploratory studies}
In this section, we evaluate the impact of several parameters on the performance of the proposed metric to further understand their effect. 
In particular, we first analyze how the support region sizes influence the performance of individual predictors and total quality scores. 
Secondly, we explore the usage of different color spaces for the definition of textural descriptors. 
Thirdly, we study the effect of using predictors coming from only one out of the two attribute domains, namely, geometry or texture, to validate our selection of both. 
Lastly, we investigate the usage of different regression models to fuse individual predictors to total quality scores.

\subsection{Support regions}
In this first study, we aim to understand the impact of varying the size of the support regions over which we compute descriptors and statistical features on the performance of our metric. 
Please note that for the former, we use $r$-search to define a support region, whereas for the latter we employ $k$-nn, as explained in sections~\ref{ssec:descriptors} and~\ref{ssec:statistical_features}, respectively.

It is worth noting that there is an inter-dependency between support regions for descriptors and for statistical features. 
For example, decreasing the descriptors' support region leads to descriptor values being more susceptible to noise; thus, neighboring descriptor values will exhibit greater differences, better capturing high-frequency components.
On the contrary, increasing the descriptors' support region causes a loss of fine details and is equivalent to smoothening the surface properties or applying a low-pass filter; in this case, neighboring descriptor values will be similar. 
At the same time, lowering the statistical features' support region implies that the descriptor values under consideration will be similar given that they are adjacent and reflect surface properties from very close vicinities.
Conversely, increasing a statistical features' support region decreases the error due to the larger sample size; yet, it increases the dispersion between descriptor values due to the recruitment of remote, spatially irrelevant samples.
Thus, there is a need to evaluate the effect of their configuration on the performance of the proposed metric. 
For this purpose, we initially fix the descriptors' 
and alter the 
statistical features' support region size;
then, we fix the statistical features' 
and alter the descriptors' support region size. 

\begin{figure}[!t]
\centering
\subfloat[D1]{\includegraphics[width=0.49\textwidth]{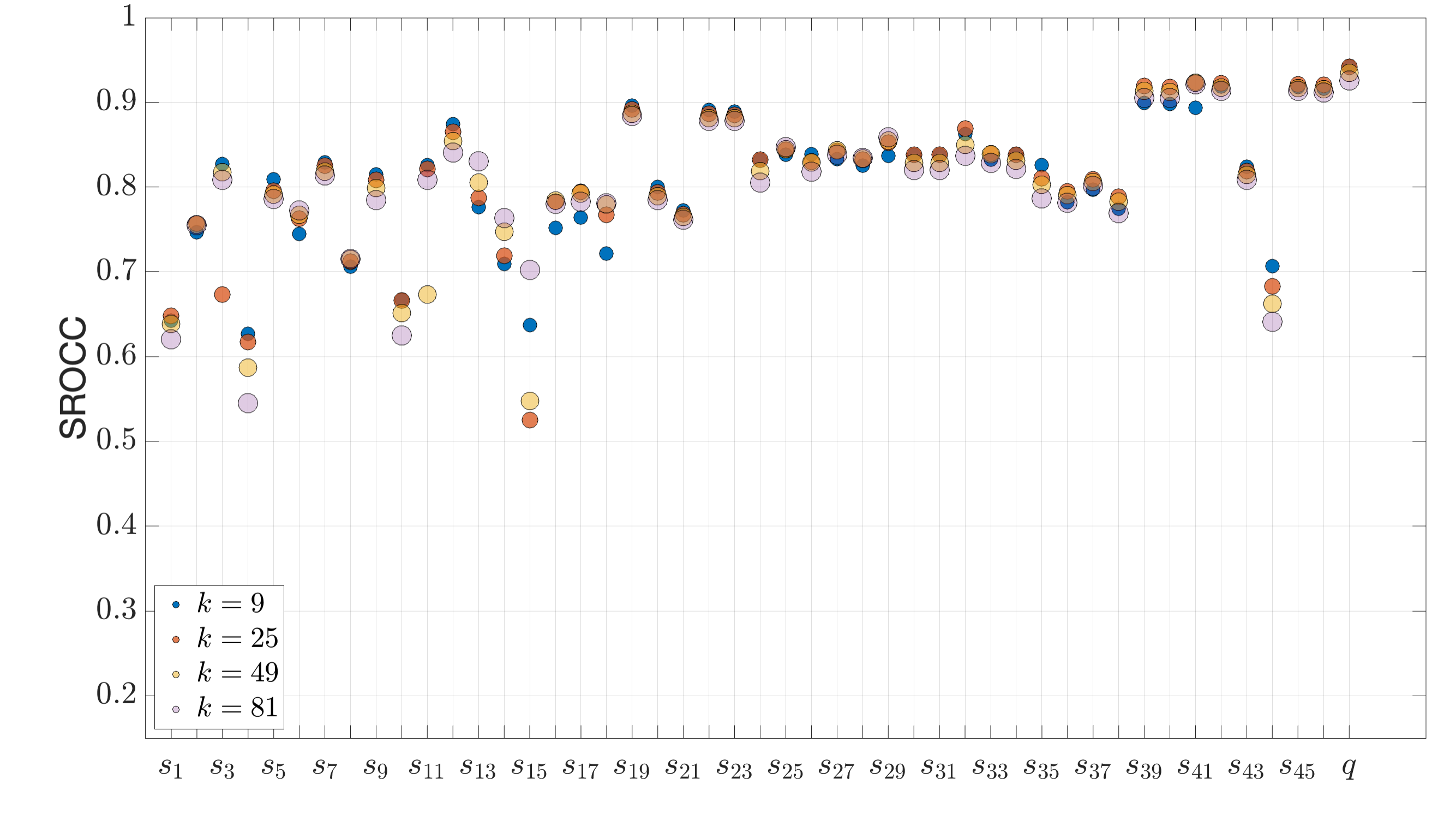}%
\label{fig:D1featsize}}
\vspace{-0.25em}
\subfloat[D2]{\includegraphics[width=0.49\textwidth]{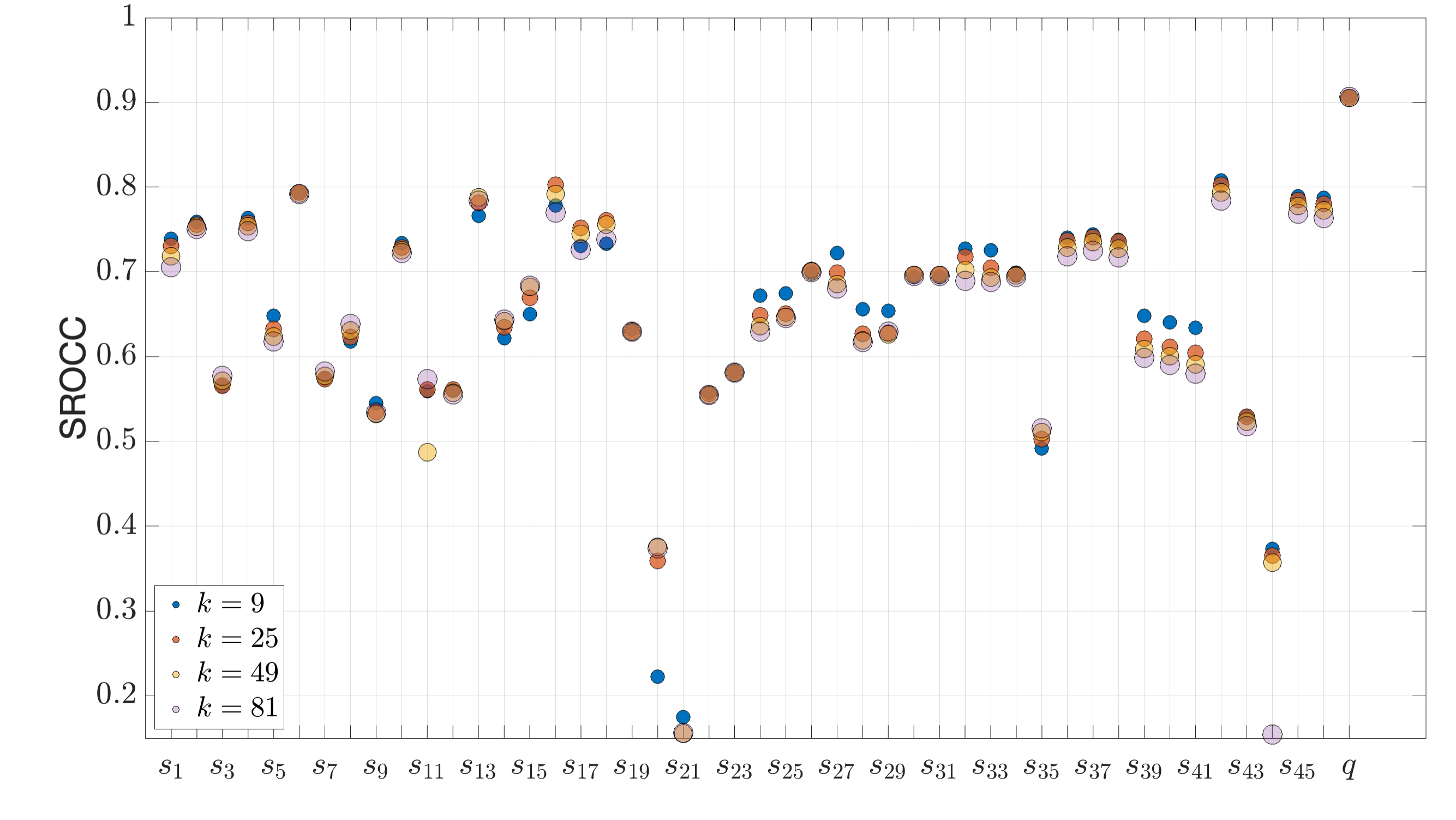}%
\label{fig:D2featsize}}
\vspace{-0.25em}
\subfloat[D3]{\includegraphics[width=0.49\textwidth]{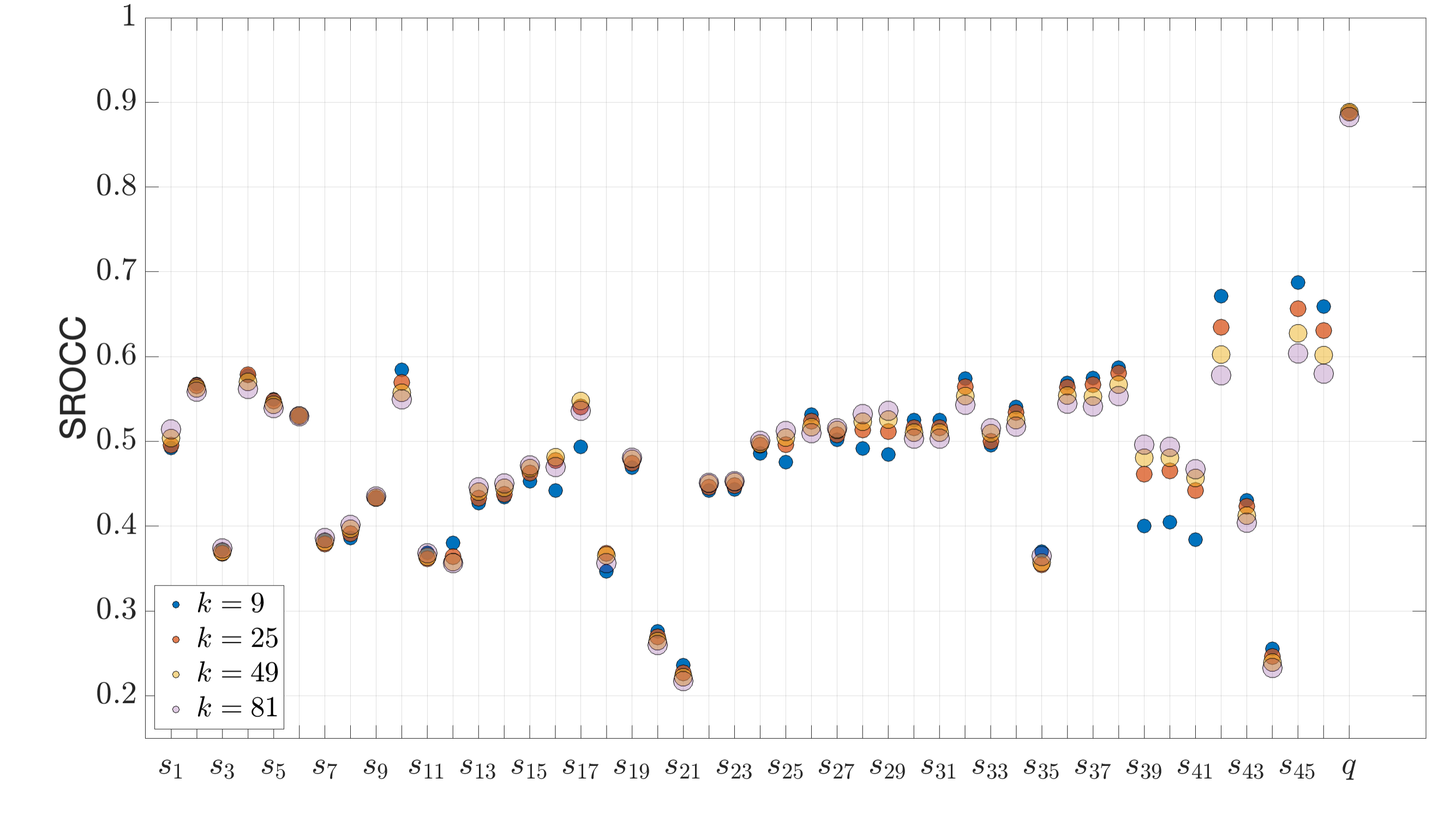}%
\label{fig:D3featsize}}
\hfil
\caption{SROCC for every predictor $s_{j}$ and average SROCC for the total quality score $q$ in every dataset, under different neighborhood sizes using the $k$-nn algorithm with $k = \lbrace 9, 25, 49, 81 \rbrace$ to compute statistical features, and the $r$-search with $r = 0.008 \times B_{R}$ to compute descriptors.}
\label{fig:featsize}
\end{figure}

\subsubsection{Support regions for statistical features}
In this case, we compute the statistical features using the $k$-nn algorithm with $k = \lbrace 9, 25, 49, 81 \rbrace$, and the descriptors using the $r$-search with $r = 0.008 \times B_{R}$.
Our selection of $k$ values is based on the fact that the point clouds of the datasets under consideration are voxelized, dense, and represent large models; thus, we may assume that small point neighborhoods represent local regions, which in turn can be approximated by planar surfaces.
The selected $k$ values represent the number of vertices in fully-occupied planes of length size equal to 2, 4, 6, and 8 times the distance between two voxels. 
Figure~\ref{fig:featsize} illustrates the \acrshort{srocc} values achieved by every predictor $s_{j}$, $1 \leq j \leq 46$ and the average \acrshort{srocc} values across all testing partitions attained by the total quality score $q$ (i.e., the PointPCA metric), with different colors indicating the performance over different $k$ values.
Recall that predictors $s_{1-23}$ make use of the mean, while predictors $s_{24-46}$ employ the standard deviation.
Moreover, for every statistic, the first 15 predictors refer to the geometry and the last 8 to the texture domain.
Our results show that the selected neighborhood size for the computation of statistical features does not have a large impact on the performance, with the trends indicating that predictors perform better under smaller rather than larger neighborhoods. 
Moreover, it can be observed that the total quality scores $q$ always outperform each individual predictor $s_{j}$.
Finally, different neighborhood sizes lead to minor differences in the performance of total quality scores, slightly favoring smaller neighborhoods.

\begin{figure}[!t]
\centering
\subfloat[D1]{\includegraphics[width=0.49\textwidth]{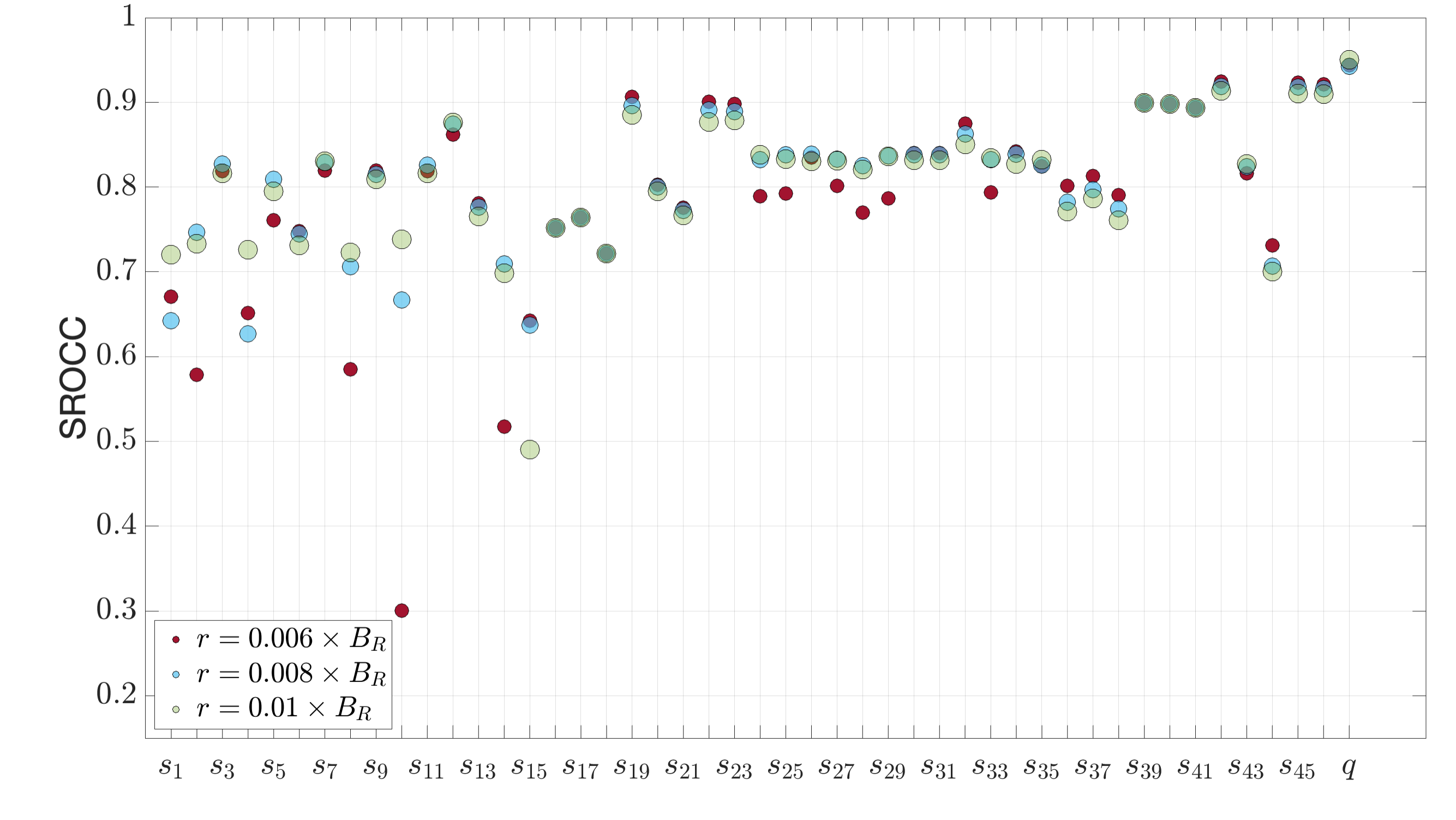}%
\label{fig:D1descsize}}
\vspace{-0.25em}
\subfloat[D2]{\includegraphics[width=0.49\textwidth]{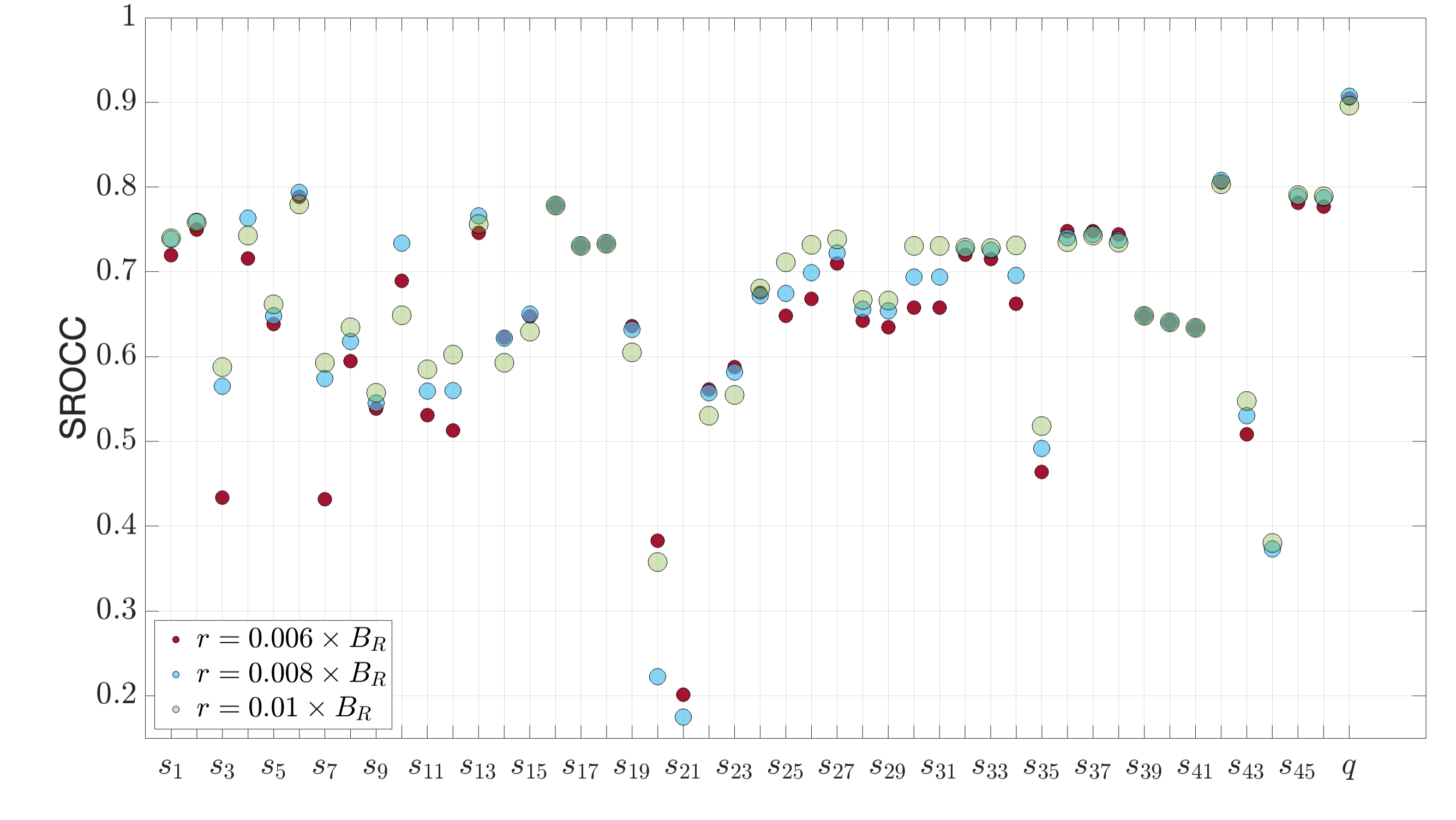}%
\label{fig:D2descsize}}
\vspace{-0.25em}
\subfloat[D3]{\includegraphics[width=0.49\textwidth]{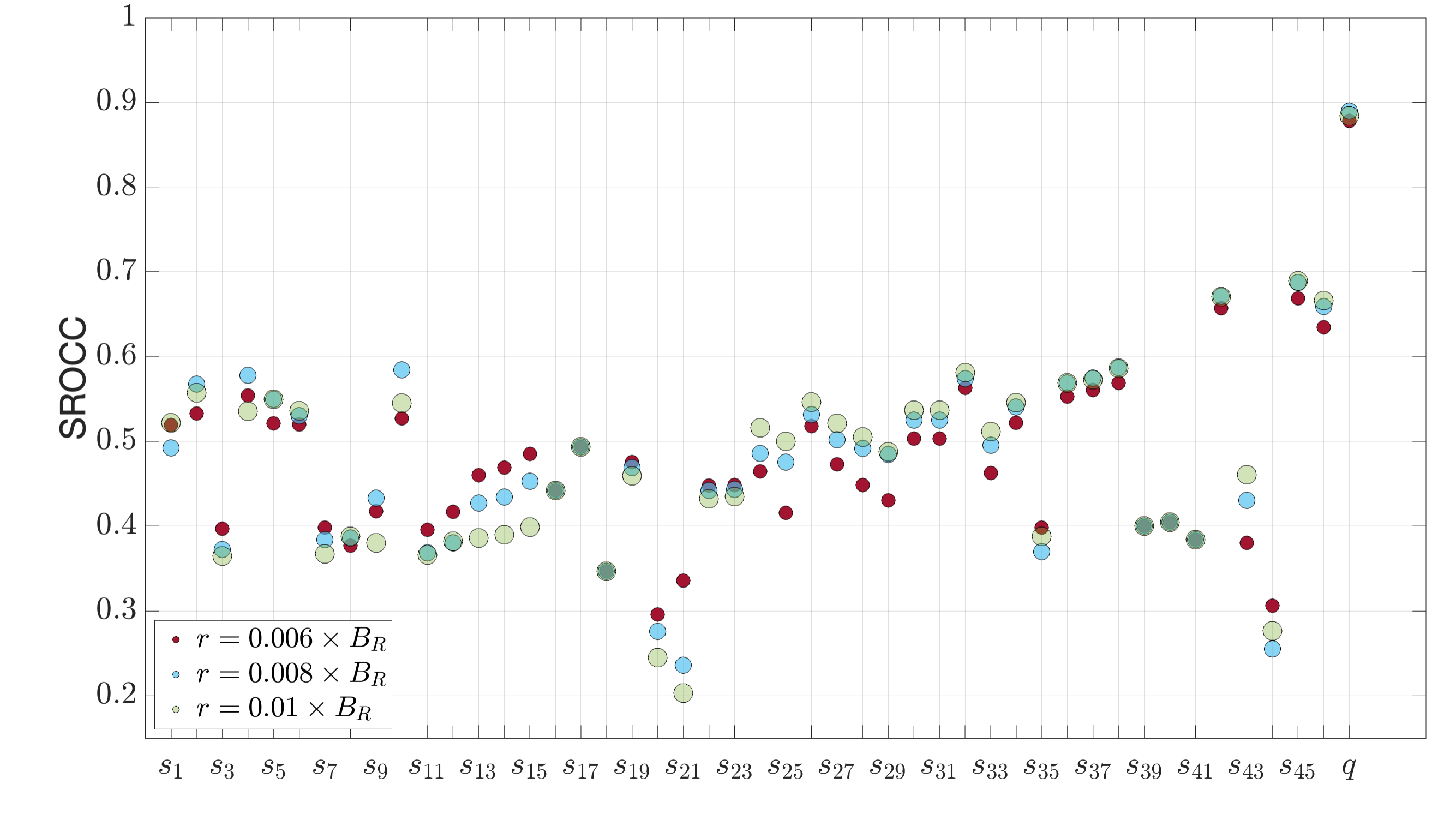}%
\label{fig:D3descsize}}
\hfil
\caption{SROCC for every predictor $s_{j}$ and average SROCC for the total quality score $q$ in every dataset, under different neighborhood sizes using the $r$-search algorithm with $r = \lbrace 0.006 \times B_{R}, 0.008 \times B_{R}, 0.01 \times B_{R} \rbrace$ to compute descriptors, and the $k$-nn with $k=9$ to compute statistical features.}
\label{fig:descsize}
\end{figure}

\subsubsection{Support regions for descriptors}
In this case, we compute the descriptors using the $r$-search algorithm with $r = \lbrace 0.006 \times B_{R}, 0.008 \times B_{R}, 0.01 \times B_{R} \rbrace$, and the statistical features using the $k$-nn with $k = 9$.
Our selection of $r$ values is inspired by the current literature (e.g.,~\cite{Alexiou2020a, Meynet2020a}), where similar volume sizes have been used to compute point cloud features for objective quality assessment.
Figure~\ref{fig:descsize} shows the SROCC values achieved by every predictor $s_{j}$ with $1 \leq j \leq 46$ and the average SROCC values of the total quality score $q$.
Our results indicate no clear pattern in the performance of mean-based predictors across all datasets, with geometric predictors (i.e., $s_{1-15}$) showing no consistent trends, and textural predictors (i.e., $s_{16-23}$) performing better in smaller neighborhoods.
For the majority of predictors that employ standard deviation (i.e., $s_{24-46}$), though, larger neighborhood sizes are preferable.
Please note that no differences can be observed across different $r$ values for textural predictors $s_{16-18}$ and $s_{39-42}$ since they do not employ a support region for the computation of corresponding descriptors (i.e., these are the non-\acrshort{pca}-based descriptors, $d^{t}_{1-3}$, equal to the \acrshort{rgb} color values).
After fusing predictors into a total quality score $q$, we observe clear benefits with respect to individual predictors $s_j$. 
Finally, considering total quality scores, marginal differences with slight gains for mid over smaller or larger neighborhood sizes are remarked.


\subsubsection{Final selection}
Our results confirm that the total quality scores lead to high prediction accuracy under all tested configurations for the descriptors' and statistical features' support region sizes. 
In the proposed settings of our metric, we set $r = 0.008 \times B_R$ and $k = 9$ for descriptors and statistical features, respectively.

\subsection{Color spaces}
In this study, we examine the performance achieved with the proposed metric by computing the same textural descriptors in alternative color spaces that are popular in the literature.
In particular, alongside the \acrshort{rgb} color space, we use the YCbCr which has been widely used for objective quality assessment; in our case, the color space conversion is performed following the ITU-R Recommendation BT.709~\cite{ITURBT7096}.
Moreover, we employ the GCM~\cite{Geusebroek2001a} which is reported to correlate well with human perception, and CIELAB~\cite{CIELAB-standard_ISO} which is recommended by the International Commission on Illumination in 1976 and designed for perceptual uniformity.
Note that in this analysis, we use all predictors from both geometric and textural domains. 
Specifically, instead of using textural predictors only, we additionally include geometric predictors to compute total quality scores, which are then compared to subjective ground truth ratings. 
This way, we don't explicitly assess the performance of the same textural predictors under different color spaces; 
but, we explore the effect of different color spaces in the performance of the proposed metric and aim to identify the one that leads to the most beneficial interactions between geometric and textural predictors.
Similarly to the analysis of section~\ref{ssec:performance_pointpca}, we learn optimal weights for all predictors per dataset and test the accuracy of the learned models in both within- and cross-dataset validation.

\begin{table*}[!t]
\caption{Performance evaluation of different color spaces (CSs). Italics indicate within-dataset results. The best performance for train/test combination across CSs is shown in bold; the second best is underlined.}
\vspace{-0.5em}
\resizebox{1\textwidth}{!}{
\centering
\renewcommand{\arraystretch}{1.2}
\label{tbl:colorSpace}
\setlength{\tabcolsep}{3.5pt}
\begin{tabular}{c l c c c| c c c| c c c}
\toprule
                          &  & \multicolumn{9}{c}{Test} \\ \cmidrule{3-11}
\multirow{2}{*}{CS}    & \multirow{2}{*}{Train} & \multicolumn{3}{c|}{D1}& \multicolumn{3}{c|}{D2}  & \multicolumn{3}{c}{D3}  \\ 
                        &        & PLCC~$\uparrow$  & SROCC~$\uparrow$ & RMSE~$\downarrow$ & PLCC~$\uparrow$  & SROCC~$\uparrow$ & RMSE~$\downarrow$ & PLCC~$\uparrow$  & SROCC~$\uparrow$ & RMSE~$\downarrow$ \\ 
\midrule
\multirow{3}{*}{YCbCr}  & D1    & \textit{0.932}	& \textit{\underline{0.939}}	& \textit{0.461}	
                        & \underline{0.795}	& \underline{0.793}	& \underline{1.466}	
                        & \textbf{0.571}	& \textbf{0.574}	& \textbf{18.794}\\
                        & D2    & 0.822	& 0.832	& 0.773	
                        & \textit{\textbf{0.935}}	& \textit{\textbf{0.911}}	& \textit{\textbf{0.838}}	
                        & \textbf{0.690}	& \textbf{0.679}	& \textbf{16.557} \\
                        & D3    & 0.775	& 0.795	& 0.856	
                        & 0.853	& 0.822	& 1.266	
                        & \textit{\textbf{0.894}}	& \textit{\underline{0.885}}	& \textit{\underline{10.142}} \\
\cmidrule{1-11}
\multirow{3}{*}{GCM}    & D1    & \textit{0.931}	& \textit{\underline{0.939}}	& \textit{0.465}	
                        & 0.791	& 0.786	& 1.477	
                        & 0.565	& \underline{0.569}	& 18.891 \\
                        & D2    & \textbf{0.828}	& \textbf{0.837}	&\textbf{0.760}	
                        & \textit{\underline{0.933}}	& \textit{0.905}	& \textit{\underline{0.856}}	
                        & 0.680	& 0.673	& 16.789 \\
                        & D3    & \textbf{0.802}	& \textbf{0.835}	& \textbf{0.810}	
                        & \underline{0.856}	& \underline{0.841}	& \underline{1.256}
                        & \textit{\underline{0.884}}	& \textit{0.878}	& \textit{10.582} \\
\cmidrule{1-11}
\multirow{3}{*}{CIELAB} & D1    & \textit{\underline{0.933}}	& \textit{\underline{0.939}}	& \textit{\underline{0.460}}	
                        & 0.775	& 0.775	& 1.524	
                        & 0.557	& 0.558	& 19.021 \\
                        & D2    & 0.821	& 0.830	& 0.774	
                        & \textit{0.931}	& \textit{0.901}	& \textit{0.869}	
                        & 0.643	& 0.632	& 17.549 \\
                        & D3    & \underline{0.794}	& \underline{0.829}	& \underline{0.825}	
                        & 0.834	& 0.825	& 1.340	& \textit{0.871}& \textit{0.864}	& \textit{11.151} \\
\cmidrule{1-11}
\multirow{3}{*}{RGB}    & D1    & \textit{\textbf{0.938}}	& \textit{\textbf{0.942}}	& \textit{\textbf{0.444}}	
                        & \textbf{0.808}	& \textbf{0.803}	& \textbf{1.423}	
                        & \underline{0.567}	& \underline{0.569}	& \underline{18.864} \\
                        & D2    & \underline{0.824}	& \underline{0.836}	& \underline{0.769}	
                        & \textit{0.932}	& \textit{\underline{0.907}}	&\textit{0.859}	
                        & \underline{0.683}	& \underline{0.678}	& \underline{16.727} \\
                        & D3    & 0.786	& 0.817	& 0.838	
                        & \textbf{0.862}	& \textbf{0.842}	& \textbf{1.229}	
                        & \textit{\textbf{0.894}}	& \textit{\textbf{0.890	}} & \textit{\textbf{10.132}} \\
\bottomrule
\end{tabular}
}
\end{table*}

In Table~\ref{tbl:colorSpace}, we present the performance indexes obtained from our metric considering different color spaces.
In general, small variations in performance can be observed. 
In the majority of cases, \acrshort{rgb} has either equivalent or marginally better performance with respect to the other color spaces. 
In particular, \acrshort{rgb} leads to better performance for within-dataset validation for D1 (PLCC = 0.938, SROCC = 0.942) and D3 (PLCC = 0.894, SROCC = 0.890), whereas it ranks second behind YCbCr for D2 (PLCC = 0.935, SROCC = 0.911 for YCbCr, against PLCC = 0.932, SROCC = 0.907 for RGB). 
For cross-dataset validation, YCbCr performs better when training on D1 and D2 and testing on D3 (training on D1, testing on D3: PLCC = 0.571, SROCC = 0.574; training on D2, testing on D3: PLCC = 0.690, SROCC = 0.679).
GCM performs better when training on D2 and D3, and testing on D1 (training on D2, testing on D1: PLCC = 0.828, SROCC = 0.837; training on D3, testing on D1: PLCC = 0.802, SROCC = 0.835).
On the other hand, RGB performs better when training on D1 and D3 and testing on D2 (training on D1, testing on D2: PLCC = 0.808, SROCC = 0.803; training on D3, testing on D2: PLCC = 0.862, SROCC = 0.842). 
However, as can be seen, the differences are rather small, showing the robustness of our metric with respect to the color space selection.

\subsection{Geometric and textural predictors}
In this study, we evaluate the impact of using predictors from different attribute domains (i.e., geometry or texture) on the proposed metric.
To do so, we compute total quality scores considering geometry-only  (i.e., $[s_{1-15}$, $s_{24-38}]$) and texture-only predictors (i.e., $[s_{16-23}$, $s_{39-46}]$), and we compare their performance with respect to using the whole set (i.e., $[s_{1-46}]$). 

Results are shown in Table~\ref{tbl:adSpace} for all datasets. 
It can be observed that for within-dataset validation, using both attribute domains leads to steadily better performance with respect to only using one. 
For D1 and D3, using textural information only leads to better performance with respect to using geometry only (D1: PLCC = 0.930, SROCC = 0.941 for texture only, versus PLCC = 0.903, SROCC = 0.907 for geometry only; D3: PLCC = 0.823 SROCC = 0.812 for texture only, versus PLCC = 0.662, SROCC = 0.625 for geometry only), whereas for D2, the opposite is true (D2: PLCC = 0.911, SROCC = 0.868 for geometry only, versus PLCC = 0.882, SROCC = 0.864 for texture only).
This can be explained considering the nature of the datasets, namely, while D1 and D3 contain compression distortions where geometry and texture are simultaneously affected, D2 contains several point clouds with only geometry or only texture distortions. 

For cross-dataset validation, we can observe that when testing on D1, using texture-only descriptors leads to better performance with respect to using the whole set, 
whereas when testing on D2, using the whole set leads to consistently better results. 
When training on D1 and testing on D3, textural information leads to the best performance; however, when training on D2, using the whole set is preferable. 
In general, we see that using predictors from both attribute domains leads to higher performance, followed by texture-only predictors, with geometry-only predictors denoting the least optimal solution.

\begin{table*}[!t]
\caption{Performance evaluation of different attribute domains (ADs). The best performance for train/test combination across ADs is shown in bold; the second best is underlined.}
\vspace{-0.5em}
\resizebox{1\textwidth}{!}{
\centering
\renewcommand{\arraystretch}{1.2}
\label{tbl:adSpace}
\centering
\setlength{\tabcolsep}{3.5pt}
\begin{tabular}{c l c c c| c c c| c c c}
\toprule
                          &  & \multicolumn{9}{c}{Test} \\ \cmidrule{3-11}
\multirow{2}{*}{AD}    & \multirow{2}{*}{Train} & \multicolumn{3}{c|}{D1}& \multicolumn{3}{c|}{D2}  & \multicolumn{3}{c}{D3}  \\ 
                        &        & PLCC~$\uparrow$  & SROCC~$\uparrow$ & RMSE~$\downarrow$ & PLCC~$\uparrow$  & SROCC~$\uparrow$ & RMSE~$\downarrow$ & PLCC~$\uparrow$  & SROCC~$\uparrow$ & RMSE~$\downarrow$ \\ 
\midrule
\multirow{3}{*}{\rotatebox[origin=c]{90}{{\centering Geometry}}}   & D1    & \textit{0.903}	& \textit{0.907}	& \textit{0.561}	& \underline{0.754}	& 0.695	& \underline{1.593}	& 0.438	& 0.459	& 20.609 \\
                        & D2    & 0.636	& 0.658	& 1.044	& \textit{\underline{0.911}}	& \textit{\underline{0.868}}	& \textit{\underline{0.992}}	& 0.533	& 0.499	& 19.378 \\
                        & D3    & 0.655	& 0.667	& 1.025	& \underline{0.810}	& 0.750	& \underline{1.422}	& \textit{0.662}	& \textit{0.625}	& \textit{17.059} \\
\cmidrule{1-11}
\multirow{3}{*}{\rotatebox[origin=c]{90}{{\centering Texture}}}    & D1    & \textit{\underline{0.930}}	& \textit{\underline{0.941}}	& \textit{\underline{0.469}}	& 0.737	& \underline{0.732}	& 1.640	& \textbf{0.577}	& \textbf{0.584}	& \textbf{18.707} \\
                        & D2    & \textbf{0.845}	& \textbf{0.848}	& \textbf{0.721}	& \textit{0.882}	& \textit{0.864}	& \textit{1.113}	& \underline{0.630}	& \underline{0.614}	& \underline{17.733} \\
                        & D3    & \textbf{0.827}	& \textbf{0.834}	& \textbf{0.764}	& 0.777	& \underline{0.783}	& 1.526	& \underline{\textit{0.823}}	& \underline{\textit{0.812}}	& \underline{\textit{12.929}} \\
\cmidrule{1-11}
\multirow{3}{*}{\rotatebox[origin=c]{90}{{\centering Both}}}    & D1    & \textit{\textbf{0.938}}	& \textit{\textbf{0.942}}	& \textit{\textbf{0.444}}	& \textbf{0.808}	& \textbf{0.803}	& \textbf{1.423}	& \underline{0.567}	& \underline{0.569}	& \underline{18.864} \\
                        & D2    & \underline{0.824}	& \underline{0.836}	& \underline{0.769}	& \textit{\textbf{0.932}}	& \textit{\textbf{0.907}}	& \textit{\textbf{0.859}}	& \textbf{0.683}	& \textbf{0.678}	& \textbf{16.727} \\
                        & D3    & \underline{0.786}	& \underline{0.817}	& \underline{0.838}	& \textbf{0.862}	& \textbf{0.842}	& \textbf{1.229}	& \textit{\textbf{0.894}}	& \textit{\textbf{0.890}}	& \textit{\textbf{10.132}} \\
\bottomrule
\end{tabular}
}
\end{table*}

\subsection{Regression models}
\label{ssec:regression_models}
In this study, we evaluate the performance achieved by the proposed metric when using different regression models to fuse individual predictors to a total quality score.
Specifically, the Linear regression (R1), K-Nearest Neighbors (R2), Support Vector Regression (R3), XGBoost (R4), and Multi-Layer Perceptron (R5) are examined as alternatives to the proposed Random Forest (R6), as implemented in the scikit-learn python package~\cite{scikit2011}.
For R1-R4, we use the default parameters.
For R5, we use 3 hidden fully connected layers with 128 neurons each; the input nodes are set equal to the number of predictors (i.e., 46) and the output nodes to one; 
ReLU activation function and MSE as loss function are employed.
For R6, we use MSE as a criterion for a split. 
Moreover, our experimentation on the number of trees indicates stable performance from 50 to 350 trees; hence, we keep the default configuration with 100 trees, as mentioned in section~\ref{ssec:execution}.

\begin{table*}[!t]
\caption{Performance evaluation of different regression models (RMs). The best performance for train/test combination across RMs is shown in bold; the second best is underlined.}
\label{tbl:regModels}
\vspace{-0.5em}
\resizebox{1\textwidth}{!}{
\centering
\setlength{\tabcolsep}{3.5pt}
\renewcommand{\arraystretch}{1.4}
\begin{tabular}{l c c c c| c c c| c c c}
\toprule
                    & & \multicolumn{9}{c}{Test} \\ \cmidrule{3-11}
& & \multicolumn{3}{c|}{D1}& \multicolumn{3}{c|}{D2}  & \multicolumn{3}{c}{D3}  \\ 
RM  & Train  & PLCC~$\uparrow$  & SROCC~$\uparrow$ & RMSE~$\downarrow$ & PLCC~$\uparrow$  & SROCC~$\uparrow$ & RMSE~$\downarrow$ & PLCC~$\uparrow$  & SROCC~$\uparrow$ & RMSE~$\downarrow$ \\ 
\midrule
\multirow{3}{*}{R1} 
& D1    & \textit{0.816}	& \textit{0.796}	& \textit{0.742}	& 0.396	& 0.363	& 2.192	& 0.258	& 0.113	& 22.079 \\
& D2    & 0.730	& 0.740	& 0.896	& \textit{0.866}	& \textit{0.834}	& \textit{1.145}	& 0.410	& 0.374	& 20.815 \\
& D3    & 0.568	& 0.448	& 1.116	& 0.777	& 0.734	& 1.523	& \textit{0.875}	& \textit{0.871}	& \textit{10.969} \\
\midrule
\multirow{3}{*}{R2} 
& D1    & \textit{0.922}	& \underline{\textit{0.934}}	& \textit{0.493}	& 0.786	& 0.785	& 1.499	& \textbf{0.621}	& \textbf{0.623}	& \textbf{17.963} \\
& D2    & \underline{0.893}	& \underline{0.905}	& \underline{0.611}	& \textit{0.926}	& \textit{0.902}	& \textit{0.896}	& 0.652	& 0.643	& 17.371 \\
& D3    & 0.672	& 0.693	& 1.006	& 0.804	& 0.787	& 1.442	& \textit{0.852}	& \textit{0.835}	& \textit{11.844} \\
\midrule
\multirow{3}{*}{R3} 
& D1    & \textbf{\textit{0.941}}	& \textbf{\textit{0.942}}	& \textbf{\textit{0.431}}	& \textbf{0.813}	& 0.793	& \textbf{1.409}	& 0.492	& 0.370	& 19.953 \\
& D2    & \textbf{0.910}	& \textbf{0.922}	& \textbf{0.564}	& \underline{\textit{0.928}}	& \textit{0.902}	& \textit{0.884}	& \textbf{0.685}	& \underline{0.675}	& \textbf{16.699} \\
& D3    & \textbf{0.840}	& \textbf{0.854}	& \textbf{0.738}	& \textbf{0.862}	& \underline{0.830}	& \underline{1.230}	& \textit{0.777}	& \textit{0.760}	& \textit{14.317} \\
\midrule
\multirow{3}{*}{R4} 
& D1    & \textit{0.928}	& \textit{0.932}	& \textit{0.486}	& 0.801	& \underline{0.795}	& 1.449	& \underline{0.572}	& \underline{0.577}	& \underline{18.764} \\
& D2    & 0.794	& 0.803	& 0.820	& \textit{0.924}	& \textit{0.895}	& \textit{0.909}	& 0.618	& 0.599	& 17.973 \\
& D3    & 0.677	& 0.694	& 0.998	& 0.848	& 0.807	& 1.283	& \textit{0.882}	& \textit{0.877}	& \textit{10.703} \\
\midrule
\multirow{3}{*}{R5} 
& D1    & \textit{0.916}	& \textit{0.923}	& \textit{0.520}	& 0.784	& 0.770	& 1.496	& 0.463	& 0.464	& 20.272 \\
& D2    & 0.889	& 0.903	& 0.620	& \textbf{\textit{0.932}}	& \underline{\textit{0.906}}	& \underline{\textit{0.860}}	& 0.679	& 0.667	& 16.809 \\
& D3    & 0.753	& 0.787	& 0.895	& \underline{0.854}	& 0.828	& 1.264	& \underline{\textit{0.892}}	& \underline{\textit{0.886}}	& \underline{\textit{10.238}} \\
\midrule
\multirow{3}{*}{R6} 
& D1    & \underline{\textit{0.938}}	& \textbf{\textit{0.942}}	& \underline{\textit{0.444}}	& \underline{0.808}	& \textbf{0.803}	& \underline{1.423}	& 0.567	& 0.569	& 18.864 \\
& D2    & 0.824	& 0.836	& 0.769	& \textbf{\textit{0.932}}	& \textbf{\textit{0.907}}	& \textbf{\textit{0.859}}	& \underline{0.683}	& \textbf{0.678}	& \underline{16.727} \\
& D3    & \underline{0.786}	& \underline{0.817}	& \underline{0.838}	& \textbf{0.862}	& \textbf{0.842}	& \textbf{1.229}	& \textbf{\textit{0.894}}	& \textbf{\textit{0.890}}	& \textbf{\textit{10.132}} \\
\bottomrule
\end{tabular}
}
\end{table*}

Performance results for every quality prediction model are presented in Table~\ref{tbl:regModels}.
As can be seen, the performance remains high and stable for the majority of regression models when training and testing on the same dataset; drops are observed using R1 with D1 and D2, and also using R3 with D3.
R3 is the best-performing model in D1 (\acrshort{plcc} = 0.941, SROCC = 0.942), whereas for D2 and D3, R6 is the best for within-dataset validation (D2: \acrshort{plcc} = 0.932, SROCC = 0.907; D3: \acrshort{plcc} = 0.894, SROCC = 0.890).

Regarding the performance of the tested regression models, R1 seems to be the weakest option, with limited generalization capabilities, independently of the dataset used for training. 
For the remaining regression models, the trends are similar, although different selections lead to the best generalization results, per training dataset.
For instance, when training on D1, R3 and R6 show higher generalization capabilities on D2 (\acrshort{plcc} = 0.813, \acrshort{srocc} = 0.793 for R3, \acrshort{plcc} = 0.808, \acrshort{srocc} = 0.803 for R6), while R2 is the best for D3 (\acrshort{plcc} = 0.621, \acrshort{srocc} = 0.623).
When training on D2, R3 is the best option on D1 (\acrshort{plcc} = 0.910, \acrshort{srocc} = 0.922) and D3 (\acrshort{plcc} = 0.685, \acrshort{srocc} = 0.675), while R2 and R6 achieve second-best performances, respectively.
Finally, when training on D3, R3 obtains the best performance on D1 by large margins (\acrshort{plcc} = 0.840, \acrshort{srocc} = 0.854), whereas on D2, R6 outperforms the rest (\acrshort{plcc} = 0.862, \acrshort{srocc} = 0.842) and is closely followed by R3 (\acrshort{plcc} = 0.862, \acrshort{srocc} = 0.830).

To see whether the difference in results between different regressors had statistical significance, we ran a 2-tailed t-test on the performance indexes obtained when training and testing on the same dataset, for all regressor pairs, across all the splits. For D1, R1 had statistically significant differences with respect to all other regressors, according to all performance indexes ($p < 0.001$ for all comparisons). In terms of \acrshort{plcc}, statistical differences were found between R3 and R5 ($p = 0.0248$), and between R6 and R5 ($p = 0.0465$); analogous results were obtained in terms of \acrshort{rmse} (R3-R5: $p = 0.0055$; R3-R5: $p = 0.0183$), whereas for \acrshort{srocc}, statistical differences were only observed for R3 with respect to R5 ($p = 0.0169$). For D2, R1 was the only regressor exhibiting statistically significant differences with respect to all other regressors, according to all performance indexes (for \acrshort{plcc} and \acrshort{rmse}, $p < 0.001$ for all comparisons; for \acrshort{srocc}, R1-R3: $p = 0.0010$; R1-R4: $p = 0.0028$, $p < 0.001$ for all other comparisons). Finally, for D3, we found statistically significant differences between all the regressors under test, according to all performance metrics ($p < 0.001$ for all comparisons). The latter is to be expected due to the large number of training/testing splits, which results in high degrees of freedom for the t-test. In general, the statistical test confirms our previous observations: with the exception of linear regression, all regressors under testing have similarly high performance, which demonstrates the robustness of the predictors with respect to the choice of regression models.

In conclusion, R3 and R6 lead to quality prediction models with the highest performance and generalization capabilities. 
In particular, R3 shows slightly better performance when testing on D1, whereas R6 achieves much better results on D3; on D2, R6 is the best with R3 attaining comparable performance. Overall, statistical analysis shows that differences between different regressors are not significant for D1 and D2, except R1, which was always found to be significantly different than the other regressors.
It is worth noting that for both R3 and R6, performance indexes from within-dataset validation show improvements over state-of-the-art metrics.
Finally, all regression models excluding R1 perform better than alternative metrics in D2 and D3, while they are following closely in D1, if not preceding.

\section{Conclusion}
In this paper, we propose a point cloud objective quality metric that relies on \acrshort{pca}-based shape and appearance predictors to evaluate distortions in the geometry and color domain, respectively.
Statistical functions are applied to the descriptor values in order to capture local relationships between point samples, which are compared between a reference and a point cloud under evaluation, producing predictions of visual quality for the latter.
The proposed predictors are assessed individually, showing good overall performance, with some textural variants leading to higher accuracy consistently across all tested datasets.
To boost the performance by leveraging the predictive potential of all the proposed predictors and return a single quality score, the Random Forest regression model is employed as part of our architecture.
Alternative learning-based models are examined and evaluated, indicating that non-linear variants lead to similarly high performance.
Moreover, the selection of parameter configuration, color space, and usage of descriptors from both geometry and texture domains are justified through a series of exploratory studies.
Our results show that PointPCA outperforms existing metrics in all tested datasets. 
Considering that certain predictors are more efficient against particular types of contents and degradations, future work will focus on the identification and adoption of optimal subsets of predictors, per use case. Moreover, ensemble of regressors will be tested to increase the prediction power of our predictors.


\printglossary[title=Abbreviations, type=\acronymtype]

\section*{Declarations}

\subsection*{Availability of data and materials}

The data that support the findings of this study are available from the parties that provided the data, as cited in the document~\cite{Alexiou2019b, Yang2020b, Liu2022a}. Restrictions may apply to the availability of these data, which were used under license for the current study, and so they might not be publicly available. Data are however available from the authors upon reasonable request and with permission of the third parties. 

The software developed in this work is available at the following link: \url{https://github.com/cwi-dis/pointpca_suite}, under \href{https://github.com/cwi-dis/pointpca}{PointPCA}. 

\subsection*{Competing interests}
The authors declare that they have no competing interests.

\subsection*{Funding}
This work was partially supported through the NWO WISE grant and the European Commission Horizon Europe program, under the grant agreement 101070109, \textit{TRANSMIXR} \url{https://transmixr.eu/}. Funded by the European Union.

\subsection*{Authors' contributions}
EA provided the main idea for the work, defined the framework and theoretical basis of the metric, conducted the majority of the experiments and the experimental analysis, and drafted the manuscript. XZ aided in the running of the experiments, specifically the regression models, and with the comparison with the state of the art. IV aided in the definition of the theoretical framework and experimental analysis, as well as in the drafting of the manuscript. PC provided feedback on the idea and experimental setup and aided in the drafting of the manuscript.

\subsection*{Acknowledgements}
We thank Konstantinos Ntemos and Hermina Petric Maretic for the useful discussion around the complexity of our algorithm.
%
%
%
%
%
%
%
%

\bibliography{refs}

\end{document}